\documentclass[referee]{aa}
\usepackage{psfig}
\usepackage{graphics}
\usepackage{txfonts}
\def\lesssim{\mathrel{\hbox{\rlap{\hbox{\lower4pt\hbox{$\sim$}}}\hbox{$<$}}}}
\begin{document} 
%
%
\title{A deep cluster survey in Chandra archival data. First results}
   \author{W. Boschin
          \inst{1}
          }


   \institute{Dipartimento di Astronomia, Universit\`{a} degli Studi 
              di Trieste, via Tiepolo 11, I-34131 - Trieste \\ 
              \email{boschin@ts.astro.it}
             }
\date{Received 30 July 2002 / Accepted 19 September 2002}
%

\abstract{ I present the first results of a search for clusters of
galaxies in Chandra ACIS pointed observations at high galactic
latitude with exposure times larger than 10 ks. The survey is being
carried out using the Voronoi Tessellation and Percolation technique,
which is particularly suited for the detection and accurate
quantification of extended and/or low surface brightness emission in
X-ray imaging observations. A new catalogue of 36 cluster candidates
has been created from 5.55 deg$^{2}$ of surveyed area. Five of these
candidates have already been associated to visible enhancements of the
projected galaxy distribution in low deepness DSS-II fields and are
probably low-to moderate redshift systems. Three of the candidates
have been identified in previous ROSAT-based surveys. I show that a
significative fraction (30-40\%) of the candidate clusters are
probably intermediate to high redshift systems. In this paper I
publish the catalogue of these first candidate clusters. I also derive
the number counts of clusters and compare it with the results of deep
ROSAT-based cluster surveys.

\keywords{Cosmology: large-scale structure of Universe -- Galaxies:
clusters: general -- X-rays: galaxies: clusters} }

\maketitle

\section{Introduction}
Clusters of galaxies are very important objects in
astrophysics. Observations of such systems at different redshifts can
be used, for instance, to explore galaxy evolution in very dense
environments (e.g. Dressler et al. 1997; Jorgensen et al. 1997; Kelson
et al. 1997; Ellingson et al. 2001; Nelson et al. 2001; van Dokkum \&
Franx 2001). Clusters are also crucial probes of mass distribution at
intermediate and large scales (e.g. Nichol et al. 1992; Collins et
al. 2000; Schuecker et al. 2001; Gonzalez et al 2002). Moreover, in
the scenario of hierarchical structure formation, the estimate of
cluster abundance as a function of their mass and redshift plays a
fundamental role in constraining the cosmological parameters
$\Omega_{\mathrm{m}}$ and $\sigma_{8}$ (e.g. Bahcall \& Cen 1993;
Girardi et al. 1998; Borgani et al. 2001; Reiprich \& B\"ohringer
2002). Following a different approach, observations of the
Sunyaev-Zeldovich effect (Sunyaev \& Zeldovich 1972) in a large sample
of distant clusters can also be used for a direct measurement of their
distances, and can thus provide estimates of the Hubble's constant
(Mauskopf et al. 2000; Reese et al. 2000).\\
In any case, all these studies require large samples of clusters in a
wide range of redshift, from $z=0$ to $z=1$ and beyond. In particular,
nowadays much interest is given to the search for medium-high redshift
clusters, whose physical and statistical properties need a big effort
of investigation. Up until the present, the largest samples of distant
clusters have resulted from optical/NIR and X-ray sky
surveys. Optical/NIR cluster surveys consist in searching for
enhancements in the surface density of galaxies (e.g. Postman et
al. 1996; Nonino et al. 1999). This classical technique suffers
seriously from projection effects, sometimes leading to false
identifications, particularly for poor clusters at high redshift,
where the contrast with the background galaxy surface density is
low. Another problem of these surveys is that the cluster galaxy
richness has a poor correlation with mass (e.g.  Borgani \& Guzzo
2001). This does not assure that the selected sample contains all
clusters within a given volume and above a given mass, since the
cluster selection function is only loosely related to this fundamental
parameter, which is what model predictions are based upon. Therefore
it is difficult to use optically/NIR selected clusters for
cosmological studies. The advantage of optical surveys is the
availability of a large number of ground telescopes with high quantum
efficiency detectors. This makes completing wide area surveys an easy
and relatively quick task. In fact, recent deep optical/NIR surveys
(e.g. the Las Campanas Distant Cluster Survey) are now reaching areas
in excess of 100 deg$^{2}$ (Gonzalez et al. 2001; Nelson et al. 2002)
enabling the discovery of the richest clusters at $z\sim 1$ and the
study of the evolution of the moments of the cluster galaxy
distribution.\\
Instead, X-ray cluster surveys consist in detecting the X-ray emission
from the hot gas which fills the space between cluster galaxies. Even
if large area X-ray surveys are time intensive, their big advantage is
that clusters look sharper in the X-ray than in the optical
sky. In fact, while the optical emission is approximately only linearly
dependent on the number of cluster galaxies, the X-ray emission is
proportional to the square of the local gas density. Moreover, the
detection of the intracluster medium is an unambigous indicator of a
real cluster. For this reason the problem of spurious detection is
much less severe. But the best feature of X-ray surveys is that their
selection function can be determined to high accuracy knowing the
properties of the telescope used. Moreover, since X-ray luminosity is
a good indicator of the total cluster mass (Reiprich \& B\"ohringer
1999), we can safely make cosmological predictions for clusters of a
given mass and sensibly compare them with observed clusters of a given
luminosity. In synthesis, X-ray surveys are much easier to use than
optical/NIR ones as cosmological probes to address global cluster
properties and not only for studying single, yet interesting,
objects.\\
In the past decade these good qualities encouraged the compilation of
many X-ray cluster surveys mainly based on ROSAT data (e.g. Rosati et
al. 1995; 1998; Scharf et al. 1997; Vikhlinin et al. 1998). All these
surveys gave important contributions in determining the X-ray cluster
number density and luminosity function at lower and lower flux
limits.\\
In this paper, I present the first results of a search for galaxy
clusters in deep (exposure time T$>$10 ks) Chandra ACIS pointed
observations at high Galactic latitude. The Chandra's high spatial
resolution is the strong point of a survey based on this telescope,
since it is essential to efficiently separate between pointlike and
extended sources and, in particular, to reduce the problem of the
contamination of distant clusters by active galactic nuclei. The
survey is being carried out using the Voronoi Tessellation and
Percolation technique (VTP, Ebeling \& Wiedenmann 1993). This approach
is completely non parametric, i.e. it is equally sensitive to
spherically simmetric structures and to irregular ones, and shows its
effectiveness to detect low surface brightness sources with greater
sensitivity when compared with standard detection methods (Scharf et
al. 1997).\\
This paper is arranged as follows: in Sect. 2 I present a brief
overview of the Chandra selected pointings and the data reduction; in
Sect. 3 I describe the source detection method and the steps of the
construction of the catalogue. In Sect. 4 I present the calibration of
the survey sky coverage. Sect. 5 presents an estimate of the redshift
distribution of the detected candidates and a measure of the sky
volume probed by the survey. In Sect. 6 and 7, I present the
log$\,N$-log$\,S$ relation and the catalogue of detected
candidates. Finally, in Sect. 8, I summarize the main results and give
my conclusions.\\
Computations throughout the paper use $\Omega_{\mathrm{m}}=0.3$ and
$\Omega_{\Lambda}=0.7$. The Hubble constant is $H_{0}=75$ km s$^{-1}$
Mpc$^{-1}$, unless otherwise stated. All X-ray luminosities and fluxes
are reported in the 0.5-2.0 keV energy band.

\section{Observations and data analysis}
\begin{figure}[t]
\psfig{figure=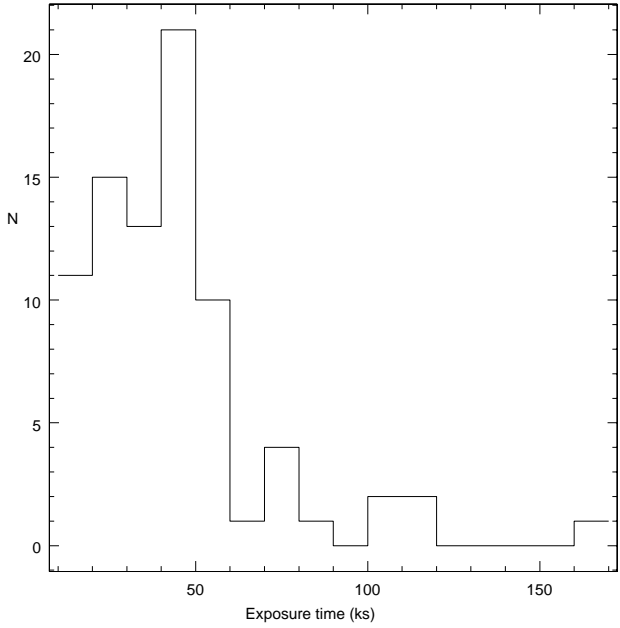,width=9cm,angle=0}
\caption{Histogram of the exposure times for the 81 selected Chandra ACIS pointings.}
\label{fig:H3892F01.ps}
\end{figure}

The data used in this paper are 81 Chandra ACIS pointed observations
downloaded from the public archive. Selected observations have
exposure times larger than 10 ks (see Fig.~\ref{fig:H3892F01.ps}) and
absolute galactic latitude larger than 20 deg.
\begin{figure}[b]
\psfig{figure=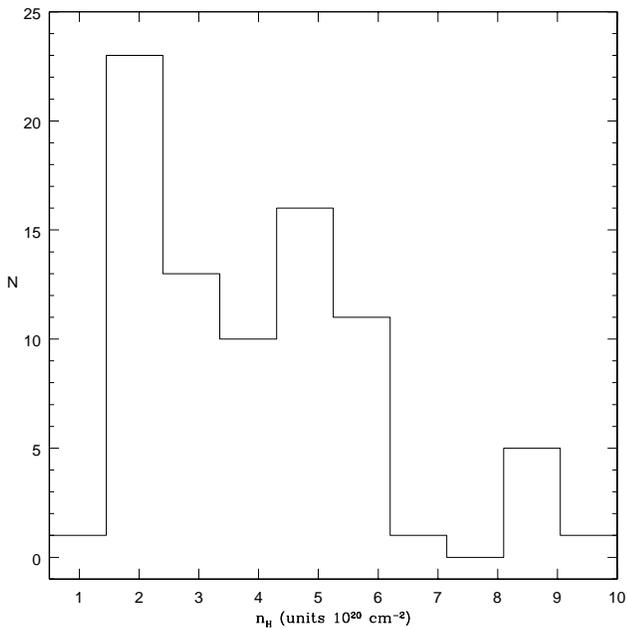,width=9cm,angle=0}
\caption{Histogram of the galactic neutral hydrogen column density
along the line of sight of the 81 selected Chandra ACIS pointings.}
\label{fig:H3892F02.ps}
\end{figure}
\begin{figure}[h]
\psfig{figure=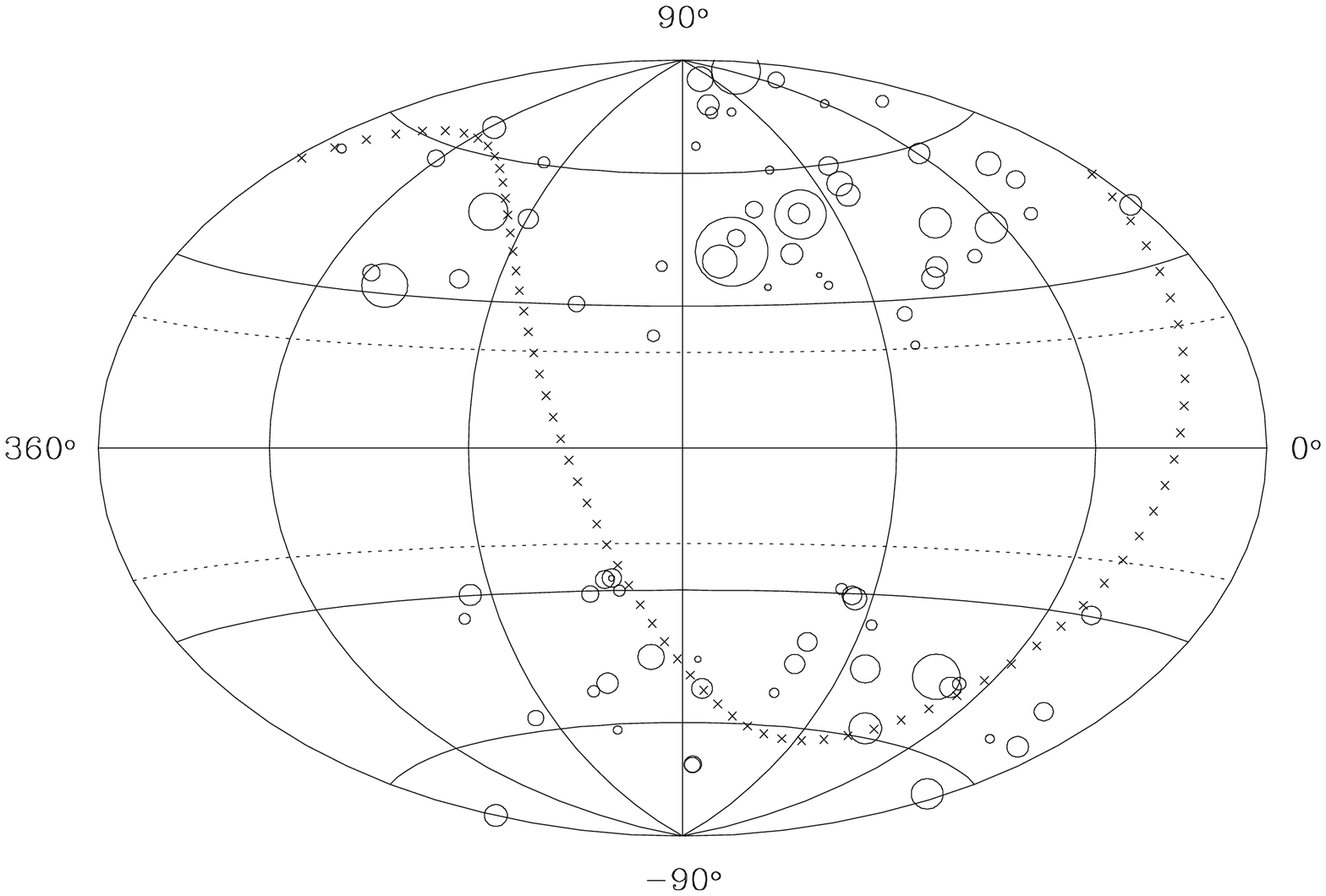,width=9cm,angle=0}
\caption{The 81 Chandra ACIS pointings selected for this survey
(Aitoff projection in galactic coordinates). The sizes of points are proportional to the exposure time. Dotted horizontal lines delimit $|b|=20^{\mathrm{o}}$; hatched
line is the celestial equator ($\delta=0^{\mathrm{o}}$).}
\label{fig:H3892F03.ps}
\end{figure}
This sky-selection criterium minimizes the effects of absorption of
extragalactic X-rays from the galactic neutral hydrogen concentrated
around the Milky Way plane. In general, the galactic hydrogen column
density along the line of sight of the selected pointings is
n$_{\mathrm{H}} < 10^{21}$ cm$^{-2}$ (see
Fig.~\ref{fig:H3892F02.ps}). I preferentially avoid pointings whose
targets are bright extended sources (typically nearby clusters and
galaxies), since they dominate the field of view and make it
impossible the detection of other objects. This also ensures that the
cluster selection function is not significantly affected by the known
angular correlation between galaxy clusters. In
Fig.~\ref{fig:H3892F03.ps}, the positions of the 81 selected Chandra
fields are plotted in Galactic Coordinates. The size of the points is
proportional to the exposure times, which ranges from $11$ to $168$
ks.\\
The Chandra ACIS is a scientific instrument made of two blocks. One,
ACIS-Imager (ACIS-I), is a square matrix of 4 1024x1024 pixels
front-illuminated (FI) CCDs (chips I0, I1, I2 and I3) covering a field
of view of about 17 by 17 arcmin and is used for imaging
observations. The other one, ACIS-Spectrometer (ACIS-S), is an array
of 6 1024x1024 pixels CCDs (4 front-illuminated, 2 back-illuminated
(BI)) covering a field of view of about 50 by 8 arcmin and is used
both for imaging observations and, mainly, for spectroscopical
observations with gratings.  In general, during an observation ACIS
can operate simultaneously both with ACIS-I CCDs and with ACIS-S CCDs,
for a maximum of 6 active CCDs. I consider here both images taken by
ACIS-I chips (chips I0, I1, I2 and I3), and ACIS-S imaging
observations (aimpoint on S3) taken by chips S2, S3, I2 and I3. In
ACIS-S observations I take into consideration only the FI chips (S2,
I2 and I3). I choose to search for X-ray sources only in these chips
since they manifest a similar photon background. This is important for
a correct use of the source detection algorithm (see
Sect.~\ref{sec:voronoi}). Moreover, in the selected ACIS-S pointings
the S3 chip is often completely filled by target photons and,
therefore, unusable for the search of other sources.\\
Before Chandra observations be used to search for sources, they need
some reduction. I perform the data reduction using the dedicated
software CIAO (Chandra Interactive Analysis of Observations, Fruscione
\& Siemiginowska 2000). For each pointing, first I filter the
preprocessed event files selecting only events with a status equal to
zero and with ASCA grades 0,2,3,4 and 6. This particular grade
selection appears to optimize the signal-to-background ratio of
sources (Chandra X-ray Center Team, 2001). Then, I clean bad offsets
intervals and examine the data on a chip by chip basis. Since for each
photon event the energy is known, I select all events with energy
between 0.3 and 10 keV. This is the band in which the energy scale of
ACIS CCDs is calibrated. For each chip I filter out bad columns and
pixels and remove times when the count rate exceeded three standard
deviations from the mean count rate per 3.3 s frame time intervals. In
fact, the background is generally quite stable during an
observation. Nevertheless, occasionally there are significant
variations, especially in the BI chips but also in FI ones. Since for
each photon event the occurance time is known, I am able to filter out
high background intervals. I then clean each chip for flickering
pixels (pixels showing unstable states of dark current signature),
i.e. I remove times where a pixel had events in two sequential 3.3 s
intervals. At the end, I remain with a clean image containing photon
events with energies between 0.3 and 10 keV.

\section{Source detection and building of the catalogue}
\label{sec:voronoi}
\begin{figure}[t]
\psfig{figure=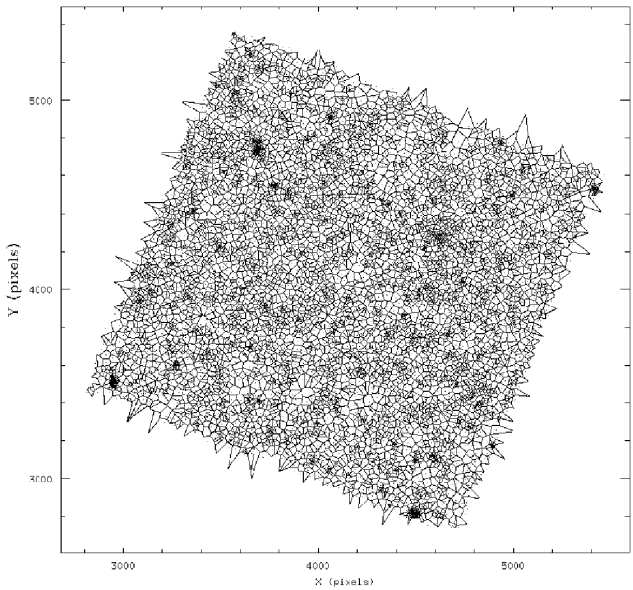,width=9cm,angle=0}
\caption{Voronoi tessellation of a typical Chandra ACIS-I photon
distribution. Photons are shown as dots and sources are immediately
apparent to the eye as clusters of small tessellation cells.}
\label{fig:H3892F04.ps}
\end{figure}

The first step in the detection process consists in the choice of an
energy band from which extract photons to pass to the detection
algorithm.  The output of the reduction step is a photon event file
containing photons with energies in a wide energy range (0.3 - 10
keV). However, it is not convenient to use the whole band in order to
detect sources. In fact, unless we are dealing with nearby massive
systems, clusters emit generally very few photons in the hard energy
band (with energies larger than 4-5 keV), therefore in order to
maximize their contrast with respect to the X-ray background and
enhance the detection probability, it is necessary to work only with
the softest X-rays. In particular, Scharf (2002) calculates the
maximum, or optimal, signal-to-noise energy band for galaxy system
X-ray emission detected by the Chandra observatory. He finds that the
``classical'' ROSAT-like 0.5-2.0 keV band is close to optimal for
clusters in a wide range of intracluster medium (ICM) temperature (in
particular for T$>$ 2 keV) and redshift $z\leq 1$. Based on these
considerations, for the detection of sources I choose to consider only
photons in the energy range 0.5-2.0 keV.\\
As a detection algorithm I use the Voronoi Tessellation and
Percolation technique (hereafter VTP). The VTP method of Ebeling
(1993, see also Ebeling \& Wiedenmann 1993) is a general method for
the detection of non Poissonian structure in a distribution of points.
By definition, a Voronoi tessellation on a two-dimensional
distribution of points (called nuclei) is a unique plane partition
into convex cells, each of them containing one, and only one, nucleus
together with the set of points which are closer to that nucleus than
to any other one (Voronoi 1908). In X-ray observations, any nucleus is
a detector pixel where at least one photon count occurs. To each of
these pixels, VTP associates a cell polygon whose sides form the
perpendicular bisectors of the non-crossing vectors joining the
nearest neighbor photon pixels. Since most cells contain only one
photon, the surface brightness associated with this photon equals the
inverse of the product of cell area and local exposure time. Multiple
counts pixels are rare and associated only to the brightest sources.
These pixels are flagged and used in the percolation algorithm
described below.\\
In Fig.~\ref{fig:H3892F04.ps}, I show the Voronoi tessellation of a
typical Chandra ACIS-I photon distribution. The sources in this field
are apparent to the eye as clusters of small cells.
After the implementation of the Voronoi tessellation, the cumulative
distribution of the inverse areas of the photon cells is computed and
compared with that expected from a random (Poisson) distribution
(Kiang 1966), and a cutoff can be determined that defines the global
background count for that field. A spatial percolation algorithm is
then run on the cells with area smaller than a given threshold above
the background, grouping them according to the excess above the
background density and forming sources. For a better background
estimation the latter two steps are performed iteratively so that, as
sources are detected, the background estimate is revised and the
source groupings are redetermined. Finally, for each source I obtain a
series of parameters: the source position (computed as a weighted sum
over photons), the detected source count rate, the detected source
area, the minimal and maxima moments of inertia of the photon
distribution, an estimate of the background count, and the probability
of the source being a statistical fluctuation. Moreover, the full set
of photons associated with each source is stored.\\
The VTP procedure is implemented on a pipeline composed by the CIAO
task vtpdetect and a C code representing an adaptation of the code
developed by Ramella et al. (2001) to detect galaxy clusters in
optical photometric surveys.

\subsection{Source deblending}
As a first step, VTP is run on each field using as a surface
brightness threshold the computed value of the background surface
brightness. However, in general it is not convenient to run VTP with a
unique surface brightness threshold. In fact, VTP has, by design, a
potentially serious drawback: it is unable to separate very close
pairs of sources, i.e, it inevitably produces ``blends'' as soon as it
finds the high-flux zones of two emission regions to be touching one
another.  In order to solve this problem, Scharf et al. (1997) show
that a practical solution is to run VTP using different surface
brightness thresholds. In this way, it is also possible to reduce the
uncertainties in source identification resulting from positive
background fluctuations that become grouped with source photons. I run
VTP three times with three different thresholds (denoted as factors of
the background surface brightness): 1.0 (the first step), 1.3 and
1.7. Then, I select the sources matching the following criteria:
\begin{itemize}
\item Sources at threshold 1.0, which do not deblend at higher
thresholds, and with an observed count rate greater than 2 $\times$
10$^{-4}$ counts s$^{-1}$ (corresponding to an observed flux limit of
$\sim$ 1 $\times \,10^{-15}$ erg s$^{-1}$ cm$^{-2}$);
\item Sources matching the above count rate criteria at threshold 1.3
that were parts of blends at 1.0 but do not deblend at threshold 1.7;
\item Sources matching the position and count rate criteria at
threshold 1.7 that were parts of blends at threshold 1.3.
\end{itemize}
\begin{figure}[t]
\psfig{figure=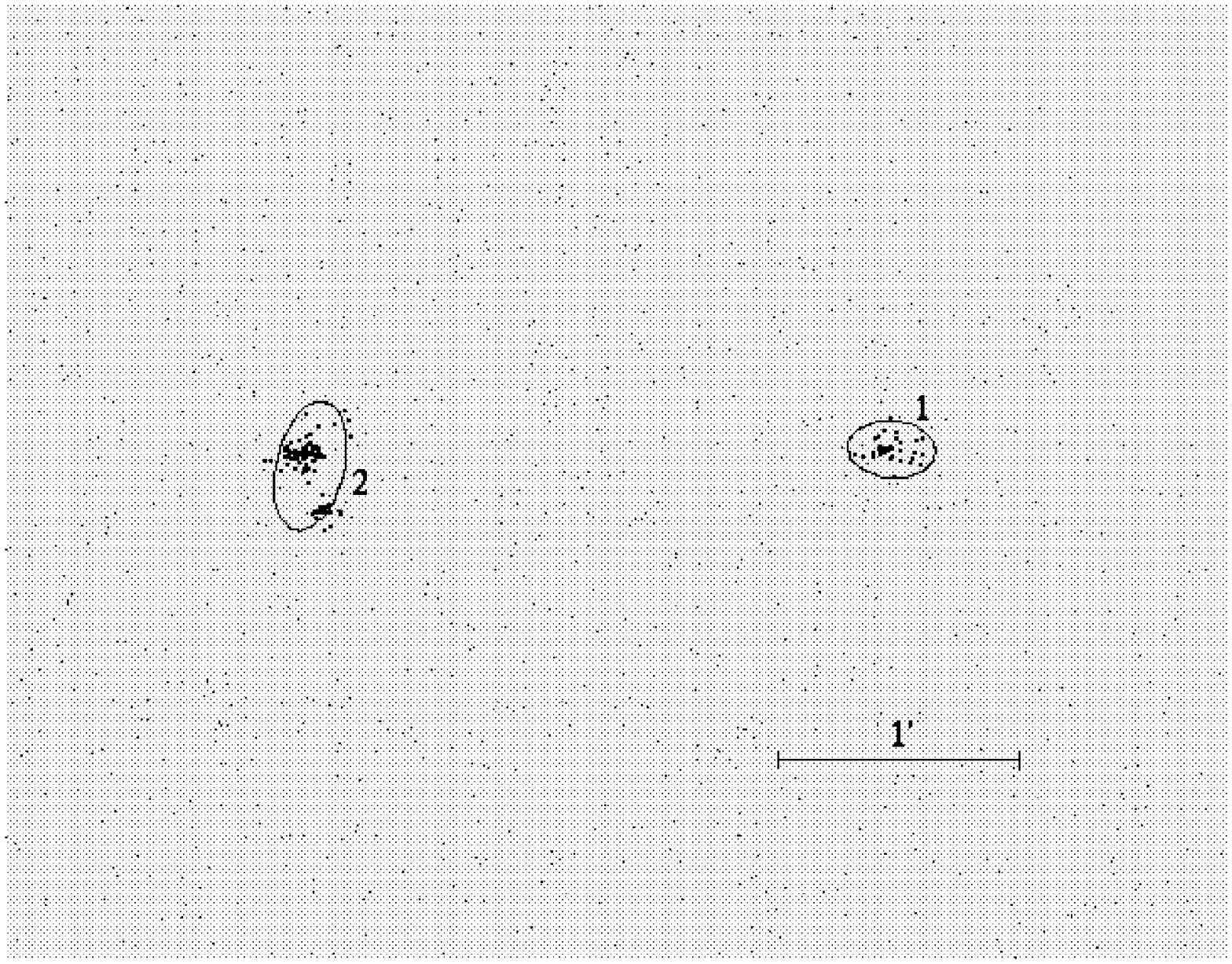,width=9cm,angle=0}
\caption{Source photons and sources identified by VTP at threshold 1.0
in a portion of an ACIS-I field. Thick points indicate VTP source
photons, sources are labeled numerically. Note that source 2 is a
clear blend.}
\label{fig:H3892F05.ps}
\vspace{5mm}
\psfig{figure=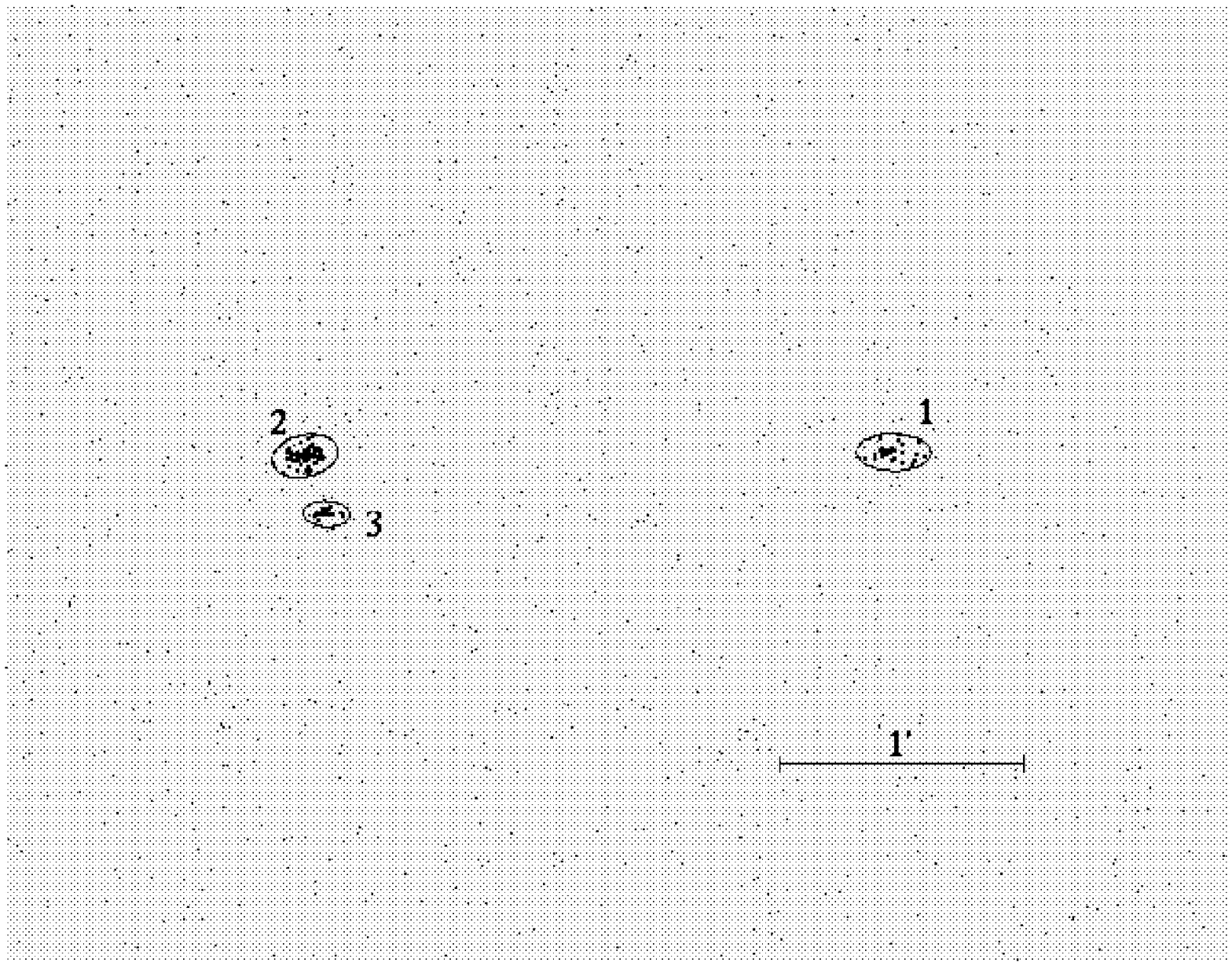,width=9cm,angle=0}
\caption{Source photons and sources identified by VTP at threshold 1.7 in the field shown in Fig.~\ref{fig:H3892F05.ps}. Note that source 2 at threshold 1.0 has now been resolved into sources 2 and 3.}
\label{fig:H3892F06.ps}
\end{figure}
This deblending procedure provides a list of candidates that is used
in a visual inspection of each field. In total, I find 3340 X-ray
sources in the 81 selected fields at the threshold 1.0. Raising the
threshold level, the number of detected sources at thresholds 1.3 and
1.7 is, respectively, 94\% and 84\% of the total number detected at
the lowest threshold. 65 sources (2\%) are deblended with the
described procedure. For some individual sources, it was necessary to
change by hand the detection threshold in order to exclude photons
that were clearly positive noise and were included in the objects
detected at threshold 1.0. These spurious photons clearly act to bias
the area estimate of the source. In general, increasing the threshold
removes the noise and typically only removes less than 10\% - 20\% of
the source photons. Moreover, these photons will be recovered in the
flux correction step described below. The threshold used for each
source is recorded and used in the correction from detected to total
flux, as detailed in Sect.~\ref{sec:correction}.\\
Figs.~\ref{fig:H3892F05.ps} and \ref{fig:H3892F06.ps} demonstrate the
differences between the results for the lowest and highest threshold
for a portion of an ACIS-I field. Photons identified as belonging to
sources by VTP are plotted in heavy type. The source numbered 2 in
Fig.~\ref{fig:H3892F05.ps} is a clear blend of two pointlike sources
when observed at the lowest threshold. In Fig.~\ref{fig:H3892F06.ps},
these two sources are deblended at the higher (1.7) threshold.\\ 
It is clear that, in principle, there may be cases of real physical
systems containing substructures that become deblended. However, a
visual inspection of the 81 fields does not reveal any case in which
deblending had split up what was probably a single extended object
into separate components. By visual inspection I also reject a set of
102 sources which appear as border effects or artifacts due to strong
X-ray sources contained in the selected fields.

\subsection{Separation between pointlike and extended sources}
\begin{figure}[!t]
\psfig{figure=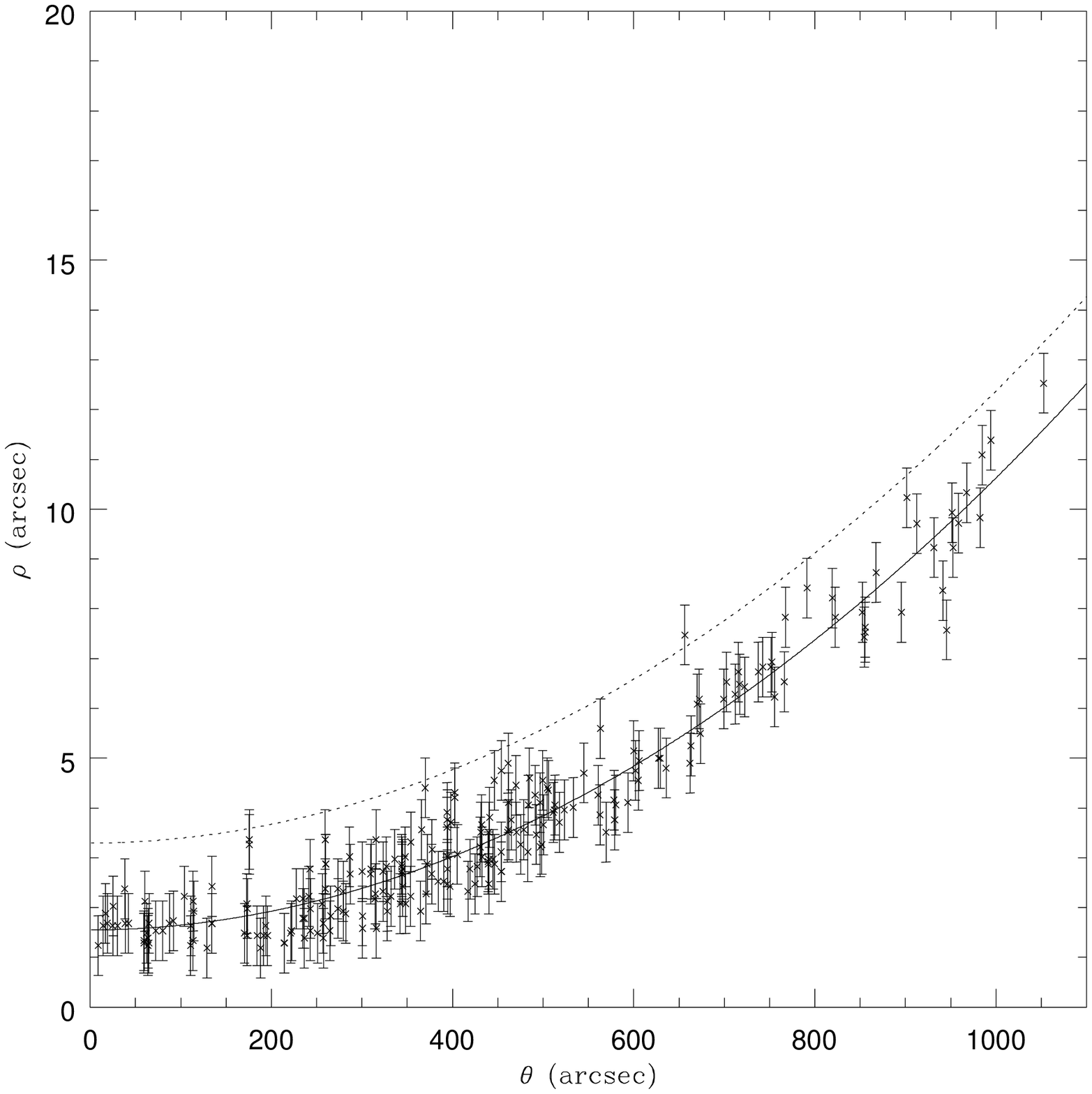,width=9cm,angle=0}
\caption{Estimated $\rho$ and 1$\,\sigma$ errors for a set of pointlike simulated sources with SNR$>$10 vs. off-axis angle. The solid line is a polynomial fit to all data points. The dotted line is a 99.9\% confidence level line derived from the $\rho$ residuals of the PSF best-fit model.}
\label{fig:H3892F07.ps}
\end{figure}

\begin{figure}[t]
\vspace{0mm}
\psfig{figure=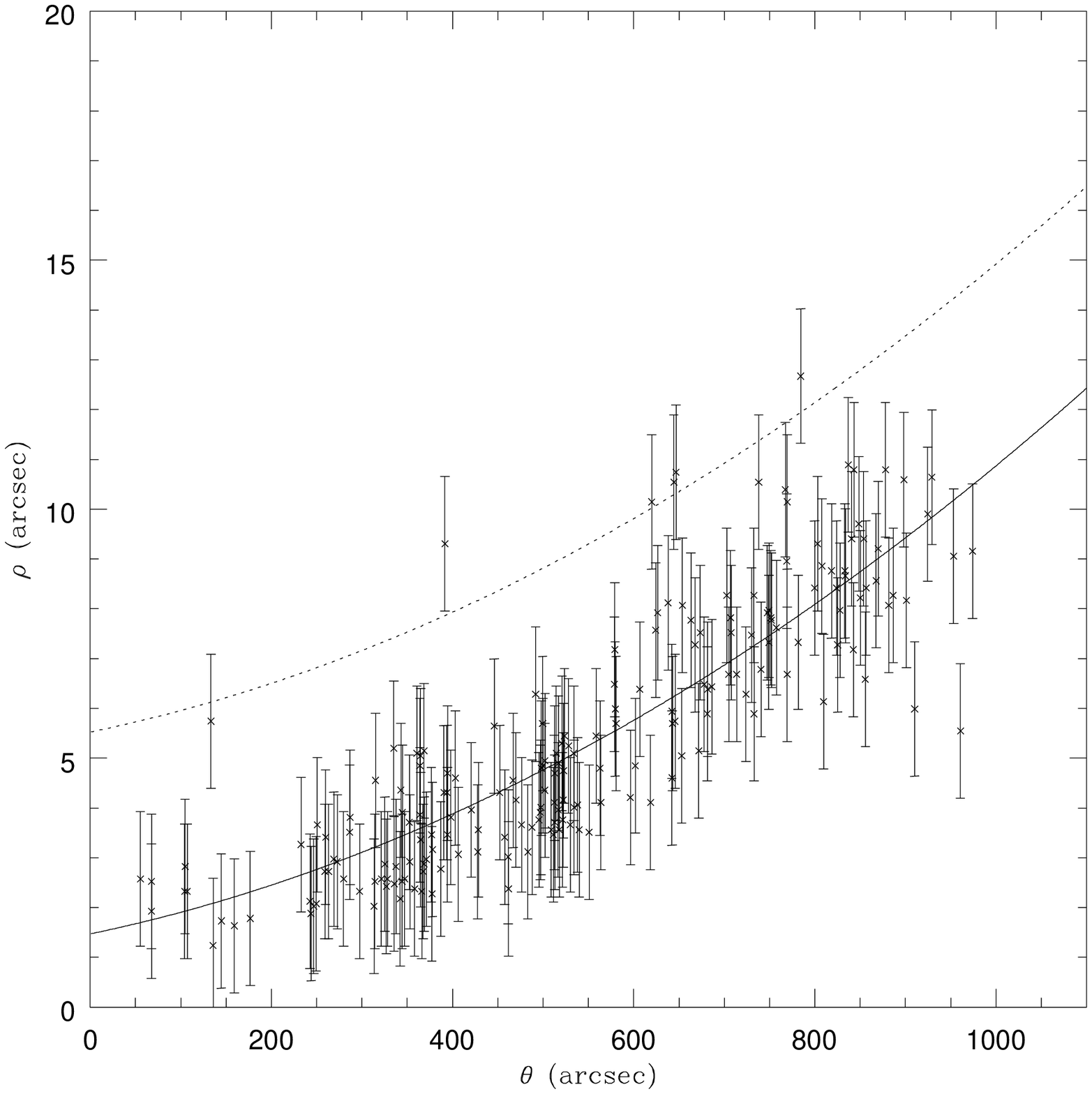,width=9cm,angle=0}
\caption{Estimated $\rho$ and 1$\,\sigma$ errors for a set of pointlike simulated sources with 6$<$SNR$<$10 vs. off-axis angle. The solid line is a polynomial fit to all data points. The dotted line is a 99.9\% confidence level line derived from the $\rho$ residuals of the PSF best-fit model.}
\label{fig:H3892F08.ps}
\end{figure}

\begin{figure}[!t]
\vspace{0mm}
\psfig{figure=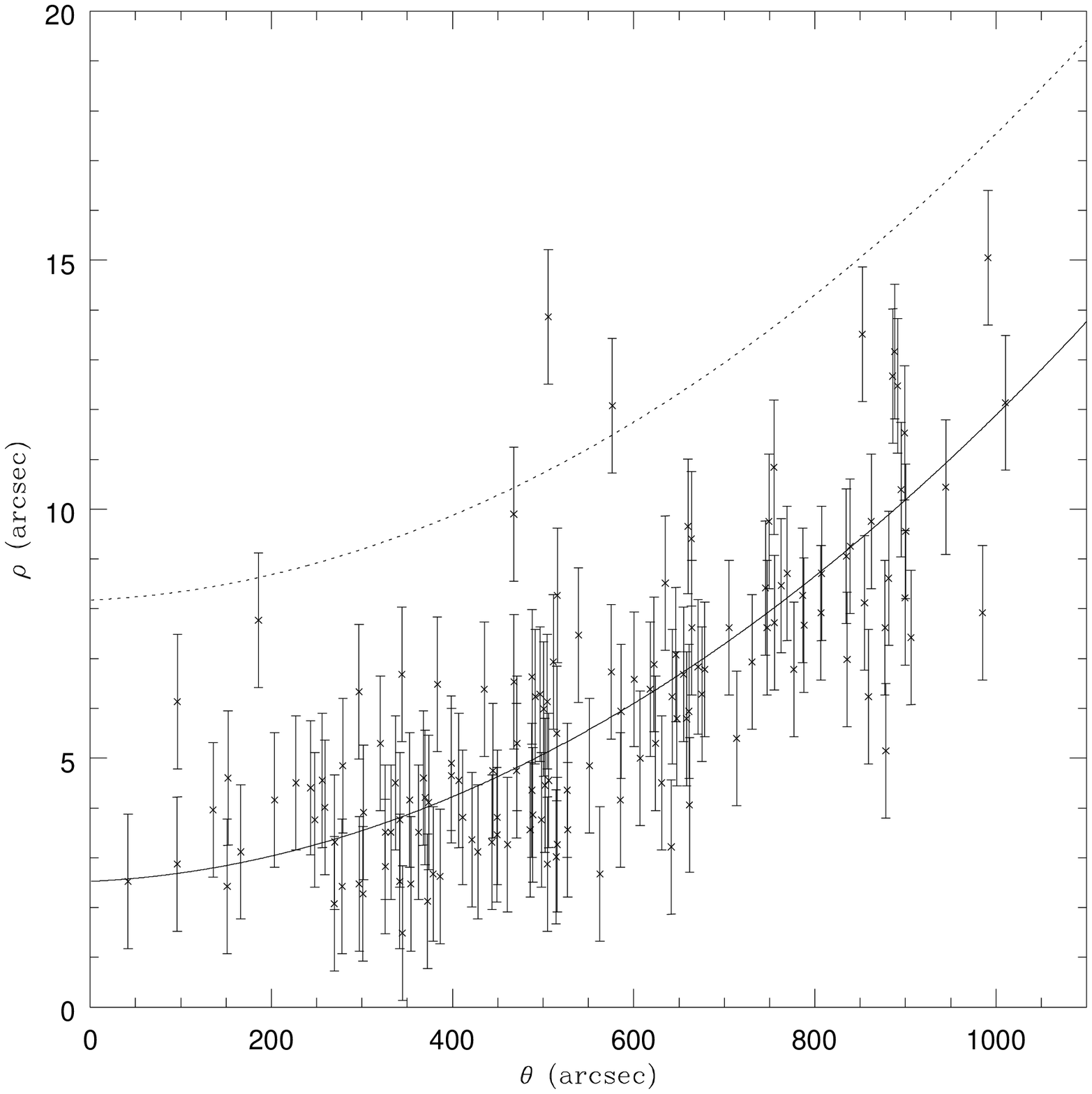,width=9cm,angle=0}
\caption{Estimated $\rho$ and 1$\,\sigma$ errors for a set of pointlike simulated sources with SNR$<$6 vs. off-axis angle. The solid line is a polynomial fit to all data points. The dotted line is a 99.9\% confidence level line derived from the $\rho$ residuals of the PSF best-fit model.}
\label{fig:H3892F09.ps}
\end{figure}

Once the deblending process is completed, the next step consists in
separating between pointlike and extended sources. Really, the very
small value of the PSF radius (less than 0.5 arcsec at the aimpoint),
makes this operation quite an easy task when considering on-axis
sources. Nevertheless, a smart procedure is necessary when considering
off-axis sources, expecially where the PSF radius is larger than about
10 arcsec. This is expecially true for sources located in chips I2 and
I3 during ACIS-S observations. In order to build such a procedure,
assumptions must be made about the nature of the sources. In
particular, I assume that sources are either intrinsically extended or
pointlike. In modeling a pointlike source I assume the PSF is
Gaussian.  The VTP algorithm returns estimates of the mean surface
brightness of the background, $\sigma_{\mathrm{bkg}}$, a background
corrected estimate $s$ of the observed count rate for each detected
source, and the area of the source above the surface brightness
threshold. From the area, I obtain an area equivalent source radius
defined as $r_{\mathrm{\scriptscriptstyle{VTP}}}\equiv
(A_{\mathrm{\scriptscriptstyle{VTP}}}/\pi)^{1/2}$. Supposing now the
source is pointlike, the detected and background corrected count rate
$s$ can be written as
\begin{equation}
s=s(r_{\mathrm{\scriptscriptstyle{VTP}}})=2\pi\,\int_{0}^{r_{\mathrm{VTP}}}\sigma_{\mathrm{\scriptscriptstyle{PSF}}}(r)\,r\,dr,
\label{eq:cap3a}
\end{equation}
where
$\sigma_{\mathrm{\scriptscriptstyle{PSF}}}(r)=\sigma_{0}\tilde{\sigma}_{\mathrm{\scriptscriptstyle{PSF}}}(r)$
is the assumed surface brightness profile for a Gaussian PSF, with
$\tilde{\sigma}_{\mathrm{\scriptscriptstyle{PSF}}}(r)=\exp(-r^{2}/2\rho^{2})$. $\rho$
is the width of the PSF.  Besides Eq. (\ref{eq:cap3a}), according to
Ebeling et al. (1996), I need a second equation specifying how close
to the background the surface brightness of the outermost regions of a
VTP detection is:
\begin{equation}
\sigma_{\mathrm{\scriptscriptstyle{PSF}}}(r_{\mathrm{\scriptscriptstyle{VTP}}})=\sigma_{0}\,\tilde{\sigma}_{\mathrm{\scriptscriptstyle{PSF}}}(r_{\mathrm{\scriptscriptstyle{VTP}}})=(f-1)\,\sigma_{\mathrm{bkg}}.
\label{eq:bkgpoint}
\end{equation}
Here $f$ is the normalized distance of the lowest surface brightness
region to the background level. This parameter is returned by VTP for
each source. Combining Eq. (\ref{eq:cap3a}) and
Eq. (\ref{eq:bkgpoint}) it is now possible to eliminate $\sigma_{0}$
and obtain an equation in the unknown $\rho$ which can be solved. I
stress that at no stage of this procedure the model profile from
Eq. (\ref{eq:cap3a}) is actually fitted to the observed surface
brightness distribution. Although I do assume a specific model
(Gaussian in this case), it is only its integral properties that
enter; thus, even if the ACIS PSF profile is not exactly a Gaussian,
deviations of the true distribution from the assumed model have much
less effect on the results than they do for the fitting procedures
employed by conventional detection algorithms. However, the lack of
any radial fitting also entails that the value for the source width
$\rho$ determined in this process cannot be expected to be as accurate
as what might be obtained in a detailed imaging analysis of pointed
data.\\
Now, the estimated $\rho$ of the detected sources can be compared with
the best-fit PSF $\rho$ as a function of the off-axis angle in order
to discriminate between pointlike and extended sources. Of course,
what happens is that extended sources will have estimated $\rho$
larger than the expected value for a pointlike source. In particular,
the lower limits of $\rho$ for a source to be considered as extended
are computed by building Monte Carlo simulations of ACIS
fields. Sources are simulated with pointlike surface brightness,
convolved with the instrumental PSF (for a nominal photon energy of
1.5 keV) and added to a representative background of Poissonian
noise. For this purpose I use the package MARX (e.g. Wise 1997), a set
of executables designed to be run in sequence to produce a
Chandra/ACIS simulation. In particular, I simulate more than 500
pointlike sources with different count rates and signal-to-noise
ratios SNR (as defined in Sect.~\ref{sec:skycoverage}) and located at
different off-axis angles. Figs.~\ref{fig:H3892F07.ps},
\ref{fig:H3892F08.ps} and \ref{fig:H3892F09.ps} show the estimated
$\rho$ for all the simulated pointlike sources vs. off-axis angles in
the SNR bins $>$10, 6-10 and $<$6 respectively. For each figure, the
solid line is a polynomial model fit to all data points, while the
dotted line is a 99.9\% confidence level line derived from the $\rho$
residuals of the PSF best-fit model. For each source, the errorbars in
the $\rho$ estimate are $1\sigma$ errors derived from the fit. A
source is considered as extended if its $\rho$ is larger than the PSF
$\rho$ at the 99.9\% confidence level in the corresponding SNR range.
In total, from the 81 selected fields I classify 51 extended sources.

\subsection{Source count rates correction for missing flux}
\label{sec:correction}
The presence of an X-ray background always limits the emission
directly detectable by any source detection algorithm to some fraction
of the total source flux. Therefore, the raw VTP count rates of the
detected sources have to be corrected for low surface brightness
emission in the far wings of the source that has not been directly
detected. This is particularly important for extended sources. In
order to perform this correction I have to assume again a surface
brightness profile for the detected sources. In particular, for the
extended sources I adopt the King's approximation to the density
profile of an isothermal sphere (King 1962) and assume a source
profile of the form (Cavaliere \& Fusco Femiano 1976)
\begin{equation}
\sigma_{\mathrm{\scriptscriptstyle{K}}}(r)=\sigma_{0}[1+(r/r_{\mathrm{c}})^{2}]^{-3\beta+1/2},
\label{eq:kingprofile}
\end{equation}
where $\sigma_{\mathrm{\scriptscriptstyle{K}}}(r)$ is the projected
surface brightness as a function of radius, $\sigma_{0}$ is the
central surface brightness and $r_{\mathrm{c}}$ is the core
radius. I also assume $\beta=2/3$. This value of $\beta$ is a good
approximation for nearby clusters (Jones \& Forman 1984). At
medium-high redshifts, clusters are characterized by values of $\beta$
in the range 0.5-0.8 (e.g. Arnaud et al. 2002; Holden et al. 2002;
Pointecouteau et al. 2002), thus $\beta=2/3$ is intermediate between
these values. Now, from the knowledge of background surface brightness
and of the VTP source characteristics I can numerically compute
$\sigma_{0}$ and $r_{\mathrm{c}}$ (see Ebeling et al. 1996 for the
details of this procedure), and the true total source count rate can
be determined from
\begin{equation}
s_{\mathrm{true}}=2\pi \int_{0}^{\infty}\sigma_{\mathrm{\scriptscriptstyle{K}}}(r)\,r\,dr=\frac{\pi\sigma_{0}\,r_{\mathrm{c}}^{2}}{3(\beta-1/2)}.
\end{equation}
\begin{figure}[!t]
\psfig{figure=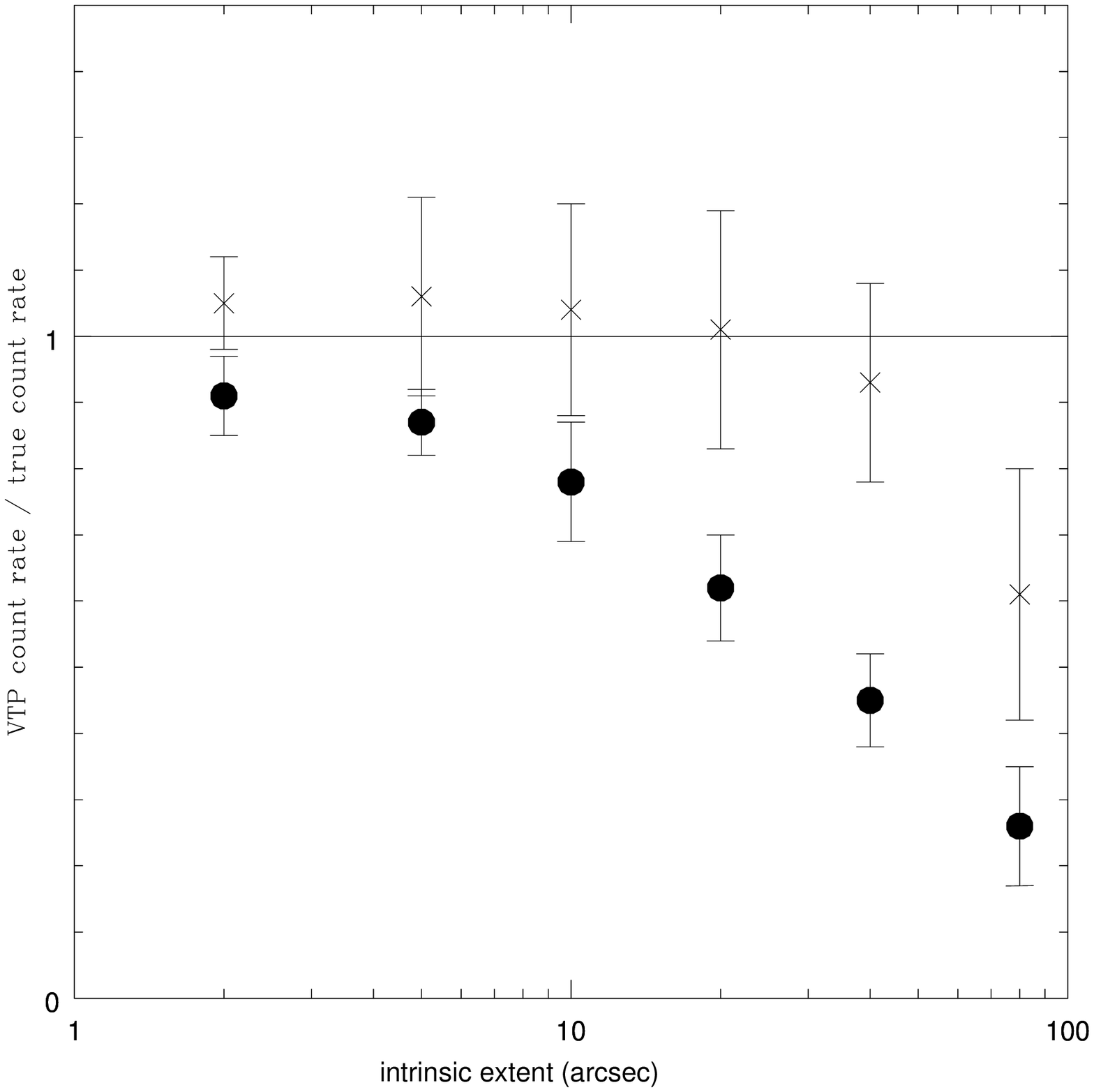,width=9cm,angle=0}
\caption{Mean ratios of detected (filled circles) and corrected (crosses) count rates to the true count rate of high SNR ($>$8) simulated sources plotted against their extent (arcsec).}
\label{fig:H3892F10.ps}
\psfig{figure=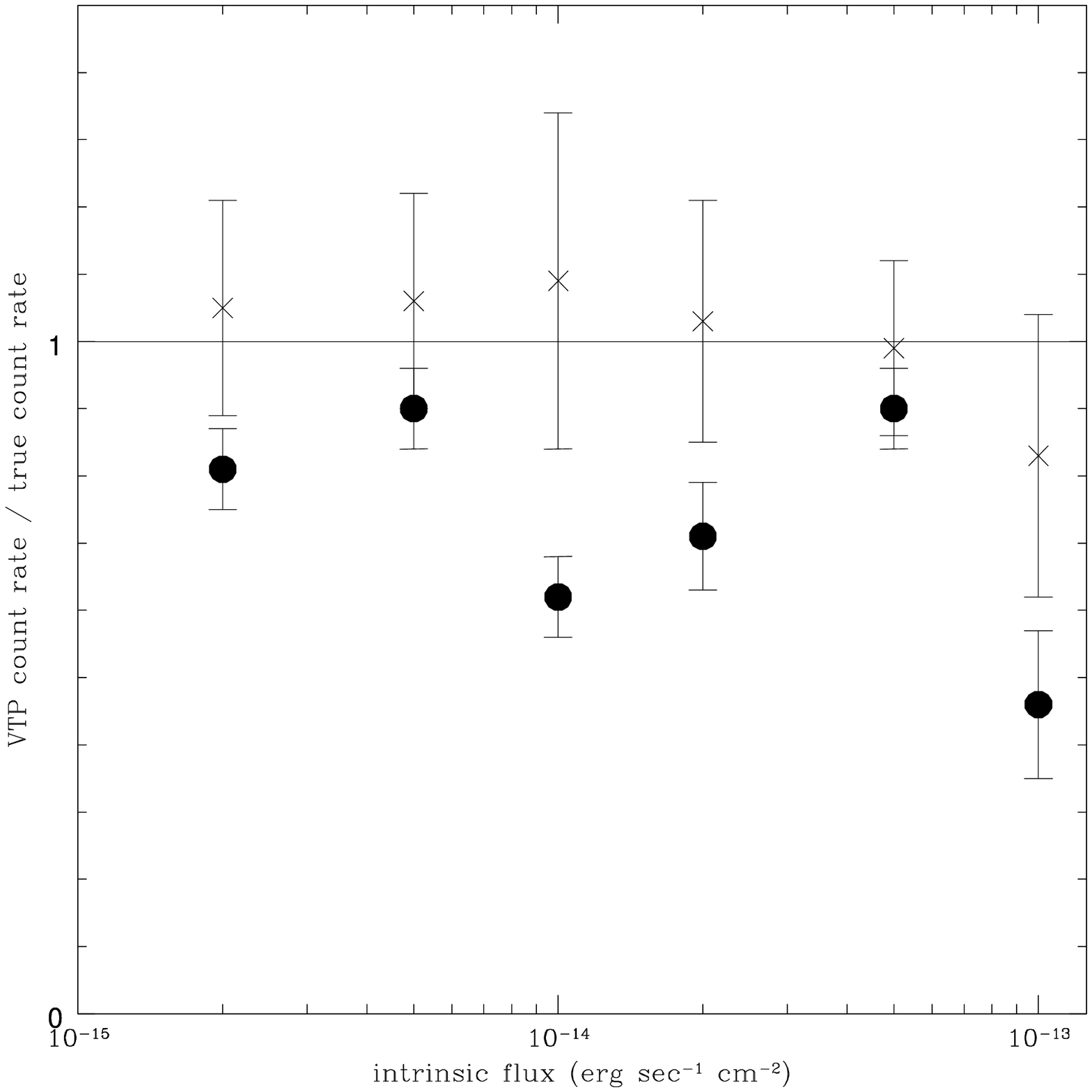,width=9cm,angle=0}
\caption{Mean ratios of detected (filled circles) and corrected (crosses) count rates to the true count rate of high SNR ($>$8) simulated sources plotted against their intrinsic flux (c.g.s. units).}
\label{fig:H3892F11.ps}
\end{figure}
In order to test this flux correction technique on the extended
sources found in ACIS fields I perform Monte Carlo simulations of ACIS
fields. By using again the package MARX I simulate more than 500
extended sources with a King profile surface brightness convolved with
the instrumental PSF. Then, the simulated sources are added to a
representative background of Poissonian noise. An ensemble of sources
are made for different sets of typical intrinsic parameters (extent,
flux, off-axis angle) to determine the expected scatter in the VTP
estimates.
\begin{figure}[t]
\psfig{figure=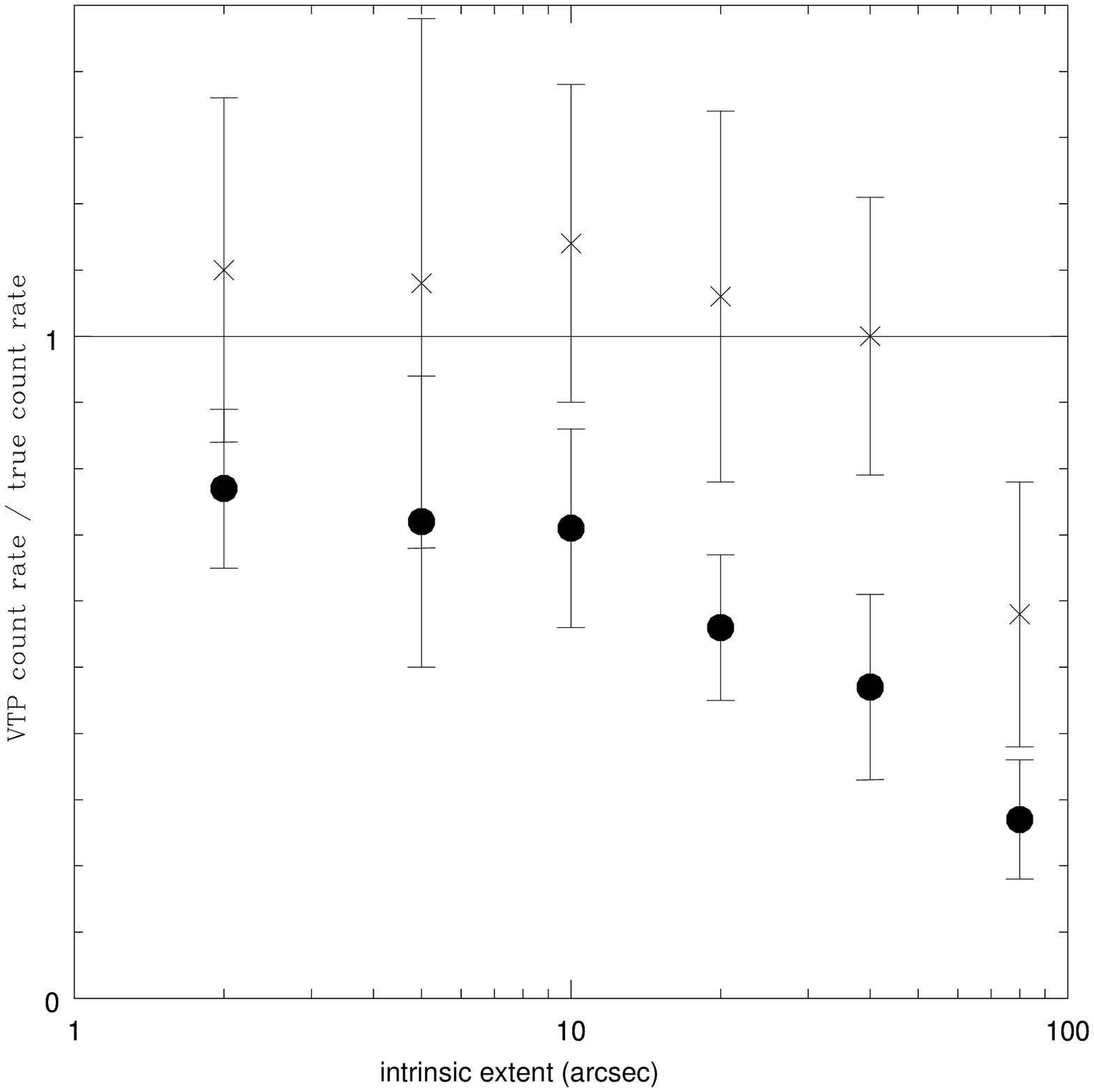,width=9cm,angle=0}
\caption{Mean ratios of detected (filled circles) and corrected (crosses) count rates to the true count rate of low SNR ($<$8) simulated sources plotted against their extent (arcsec).}
\label{fig:H3892F12.ps}
\psfig{figure=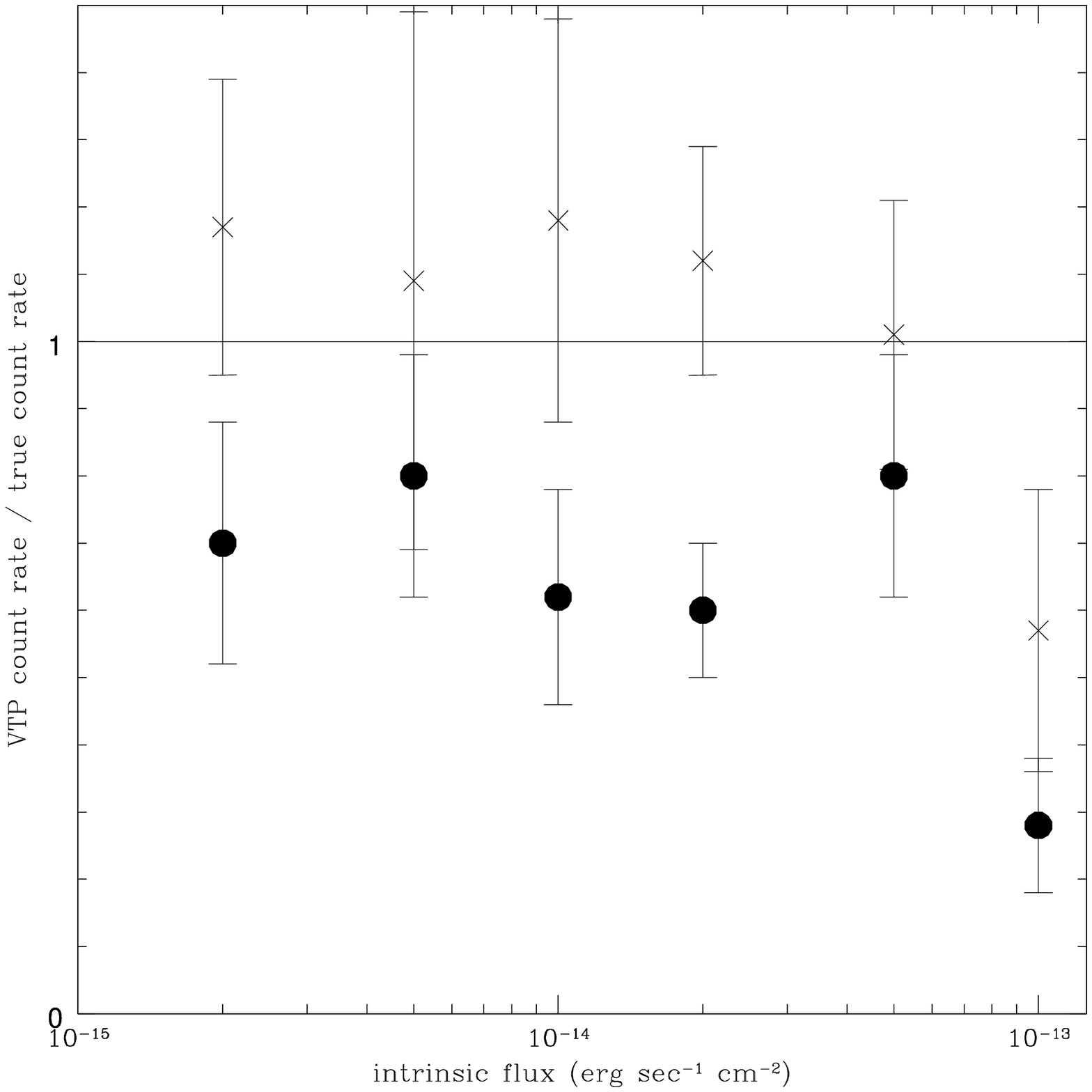,width=9cm,angle=0}
\caption{Mean ratios of detected (filled circles) and corrected (crosses) count rates to the true count rate of low SNR ($<$8) simulated sources plotted against their intrinsic flux (c.g.s. units).}
\label{fig:H3892F13.ps}
\end{figure}
In Figs.~\ref{fig:H3892F10.ps}, \ref{fig:H3892F11.ps},
\ref{fig:H3892F12.ps}, \ref{fig:H3892F13.ps} results are presented
across a large range of source parameters, from extents of $\sim 2$ to
$80$ arcsec and effective fluxes of $\sim$ 2 $\times \,10^{-15}$ to
$\sim$ 1 $\times \,10^{-13}$ erg s$^{-1}$ cm$^{-2}$. In all plots, the
ratio of the raw, detected count rates to the true count rates and the
ratio of the corrected (see above) count rates to the true count rate
are plotted on the y-axes. The effective VTP-detected S/N of a source
is a good indicator of the reliability of any flux estimate. As
expected, the efficiency of this technique to recover the true count
rate of the sources depends significatively on their SNR. Since the
median signal-to-noise ratio of the real X-ray source detection is 8,
I use this to divide the simulation results for presentation. The
effect of noise is apparent in the larger scatter seen in flux
estimates for the low S/N ($<$8) sources plotted in
Figs.~\ref{fig:H3892F12.ps} and \ref{fig:H3892F13.ps} when compared to
those of the high S/N objects in Figs.~\ref{fig:H3892F10.ps} and
\ref{fig:H3892F11.ps}. For sources of medium to small extent (i.e. $<$
40 arcsec) with S/N greater than 8, I can recover the true flux to
whitin 10\% over the flux range shown here. For sources of larger
extent (of which I might expect to see very few in this survey), the
fluxes are systematically underestimated, as expected, since most of
the flux now lies below the surface brightness limits. For the low S/N
sources (S/N $<$ 8), the same general trends are present but are
dominated by the scatter caused by noise. For small extents ($<$ 40
arcsec) I can still recover the true flux of the faintest objects to
within 10\% - 20\%.\\

\subsection{Computation of source fluxes}
\label{sec:computation}
\begin{figure}[t]
\psfig{figure=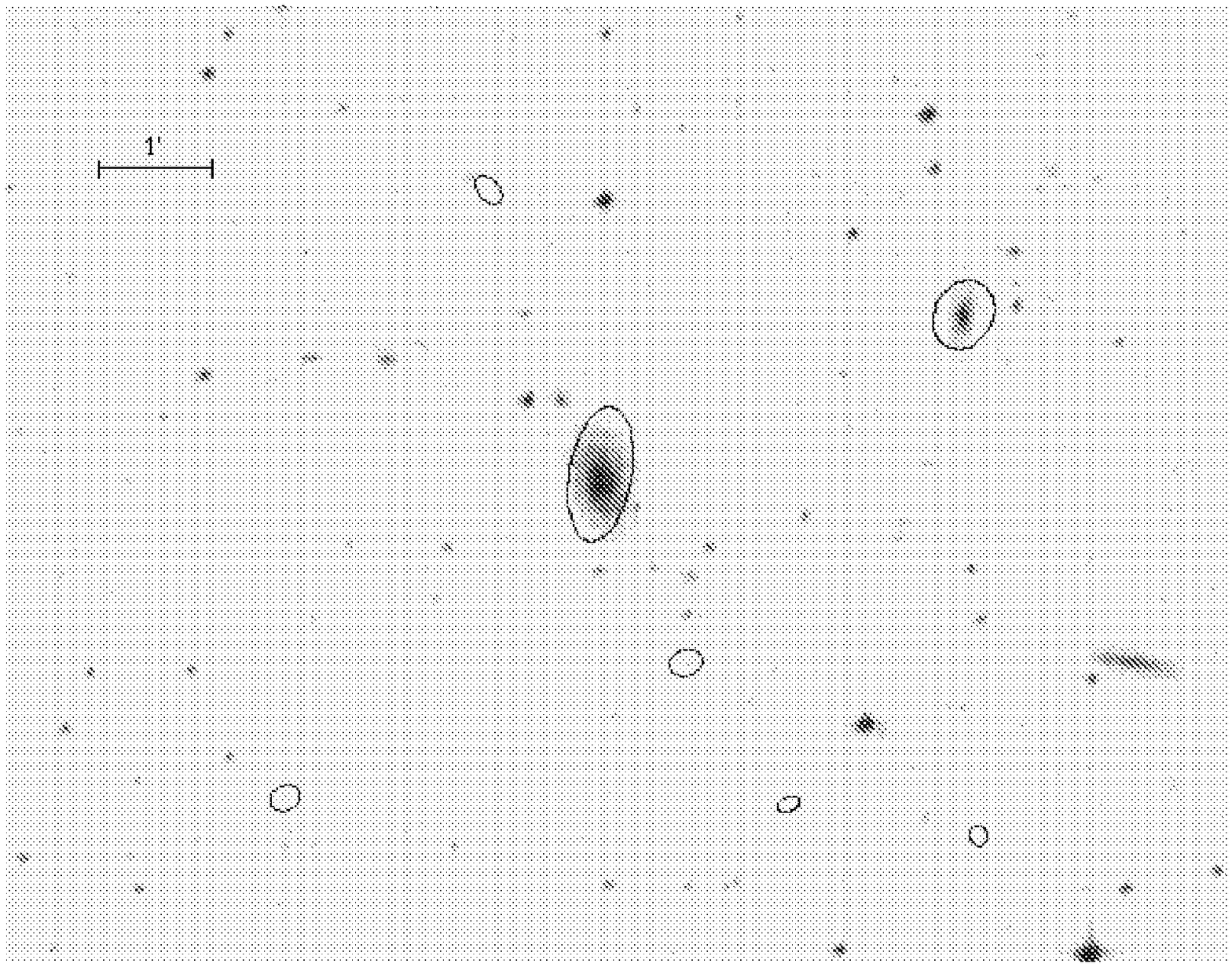,width=9cm,angle=0}
\caption{This figure represents an R-band DSS-II image of an ACIS-I
field. The ellipses represent the sources detected by VTP in the
corresponding X-ray image, two of which are clearly extended. The
image shows they correspond to relatively nearby galaxies. Therefore,
these extended sources are not listed in the catalogue of candidate
clusters.}
\label{fig:H3892F14.ps}
\end{figure}

The ACIS field experiences non uniform exposure. In ACIS-I
observations, for instance, it can vary by as much as 20\% - 25\% from
the field center to 10 arcmin radius off-axis. Therefore, exposure
maps are built for each pointing (with 4 $\times$ 4 arcsec resolution pixels)
using standard CIAO threads in order to take into account this effect
when computing source count rates.\\
After the count rate correction procedure, source count rates are
converted into fluxes multiplying them by a factor which depends on
the global characteristics of the telescope-detector system, on the
value of the line-of-sight neutral hydrogen column density, on the
energy spectrum of the source and on its redshift. I compute this
conversion factor by assuming that the extended sources exibit a
thermal Raymond-Smith (1977) spectrum with temperature $kT=5$ keV and
solar abundance ratio 0.4. For a given hydrogen column density, this
conversion factor is accurate to within 8\% for an ICM temperature in
the range 1-20 keV and a solar abundance ratio in the range 0.2-1.
The hydrogen column density along the line of sight of the sources is
taken from Dickey \& Lockman (1990). Redshifts are not known for most
of the candidate clusters at this stage of the work, therefore the
fluxes listed in the cluster catalog (see next section) should be
considered as provisional.\\
Finally, I perform a first optical follow-up of the extended sources
by analyzing the corresponding R band fields in the Second STScI
Digitized Sky Survey (DSS-II, e.g. McLean et al. 2000). These optical
fields are sufficiently deep to allow me to identify low redshift
X-ray bright galaxies (typically spiral galaxies) that appear as
extended sources in the X-ray image. Of the 51 detected extended
sources, 15 are identified as single galaxies. Therefore, these
sources are excluded from the final catalogue of candidate clusters
(as an example see Fig.~\ref{fig:H3892F14.ps}).

\section{The catalogue of candidate clusters}
\begin{figure}[!t]
\psfig{figure=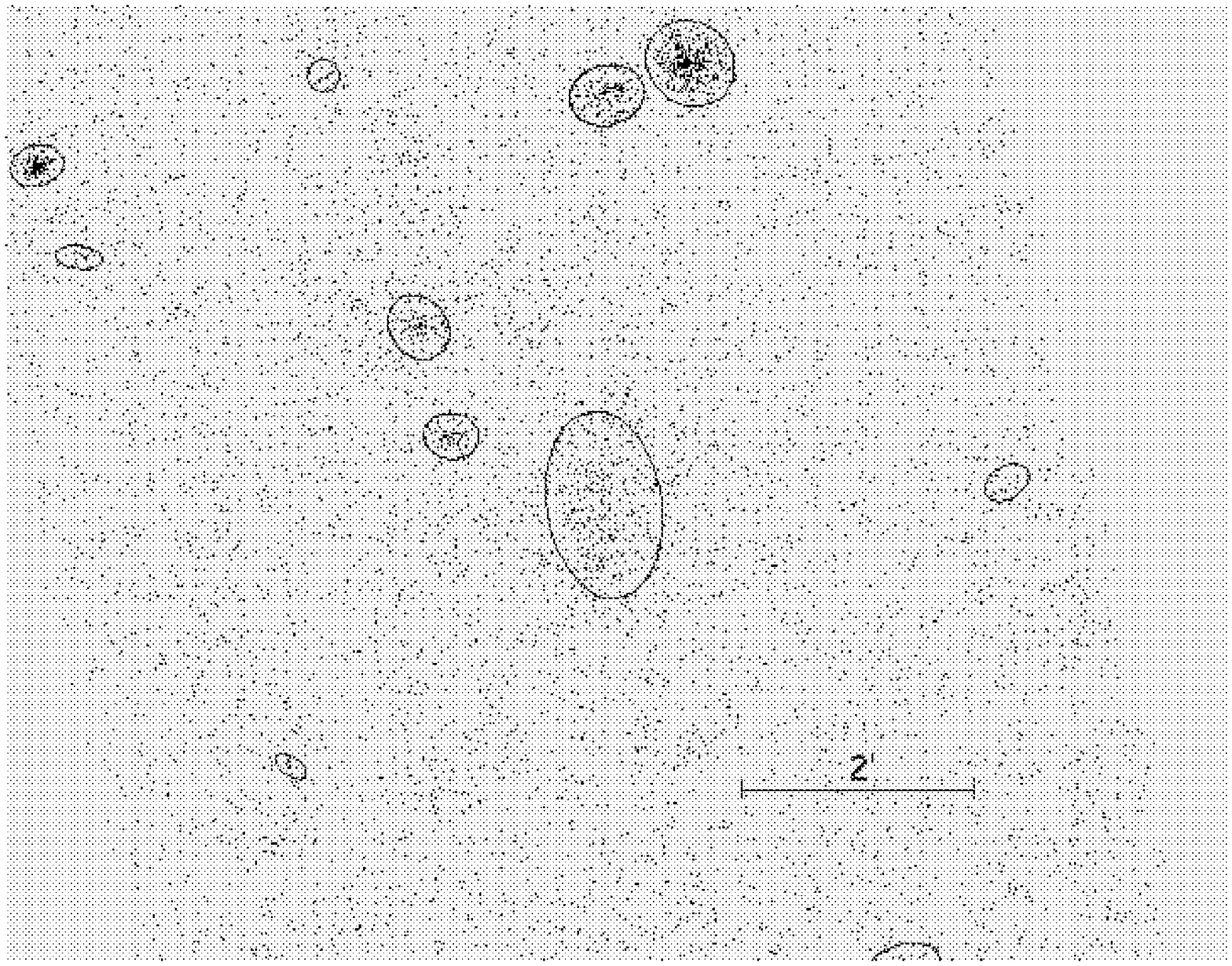,width=9cm,angle=0}
\vspace{5mm}
\psfig{figure=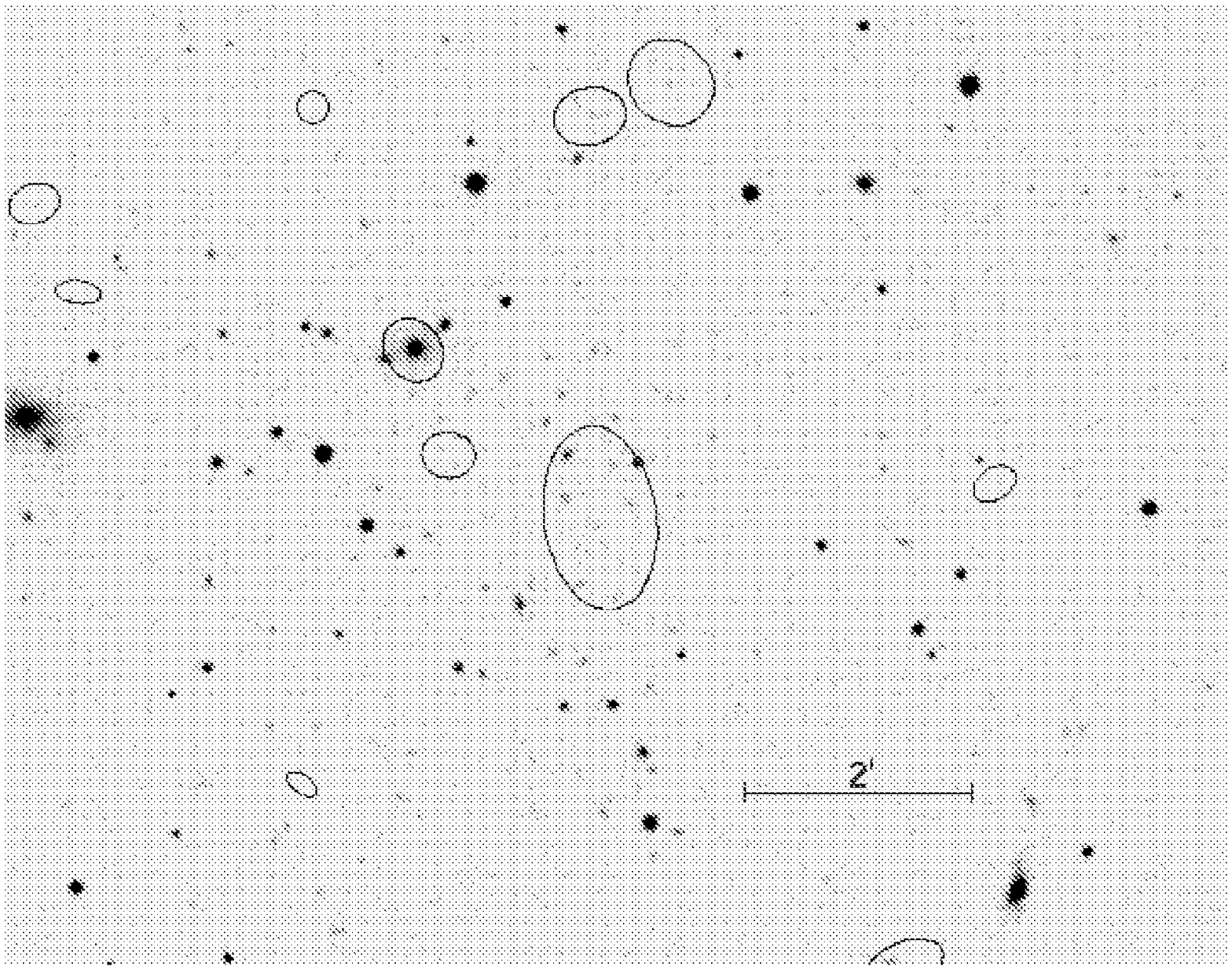,width=9cm,angle=0}
\caption{X-ray and DSS-II R band images of the ACIS-S field containing
the candidate cluster 12. Ellipses indicate sources detected by VTP in
the X-ray image. The largest one indicates the cluster. Note, in the
optical image, the presence of a concentration of faint galaxies in
correspondence with the position of the candidate cluster. It suggests
the existence of a low redshift system.}
\label{fig:H3892F15-16.ps}
\end{figure}
This survey currently covers a total sky area of 5.55 deg$^{2}$. Using
the VTP method, the sample of candidate clusters contains 36 objects
with detected flux greater than 1 $\times$ 10$^{-15}$ erg s$^{-1}$
cm$^{-2}$.  In table 1 I list the detected candidates in growing order
of right ascension. The object number is given in Col. 1. The
equatorial coordinates (J2000.0) of the X-ray centroid are listed in
Cols. 2 and 3 in degrees with four decimals (typical centroid
uncertainties are of 1-2 arcsec). The total unabsorbed X-ray flux in
c.g.s. units (erg s$^{-1}$ cm$^{-2}$) and its $1\,\sigma$ uncertainty
are listed in Cols. 4 and 5. The estimated characteristic radius
$r_{\mathrm{\scriptscriptstyle{VTP}}}$ is given in Col. 6. Col. 7
lists the signal-to-noise ratio of the sources (as defined in
Sect.~\ref{sec:skycoverage}). Finally, Col. 8 contains notes for
individual clusters. In this column, I also list the identifications
from the literature. Only five candidates seem associated in DSS-II
images to galaxy concentrations with respect to the
background/foreground galaxy distribution (see
Fig.~\ref{fig:H3892F15-16.ps}) and are probably low-redshift
systems. Three candidates are re-discoveries of systems at
low-intermediate redshift detected in previous X-ray surveys. In most
cases no optical counterpart (galaxy, group, etc.) is seen. This
suggests that most candidate clusters are probably at redshifts $z >
0.2-0.3$, thus invisible in the relatively shallow fields of the
DSS-II.
\begin{table*}[!t]
\begin{center}
\caption[ ]{The cluster catalogue}
\small
\begin{tabular}{l c c c c c r c}
\hline
Number & R.A. (J2000) & Decl. (J2000) & F$_{\mathrm{X}}$            & $\delta$F$_{\mathrm{X}}$            & $r_{\mathrm{VTP}}$ & SNR & Note\\
       &  degrees     &    degrees    &   c.g.s. units     &   c.g.s. units    &    arcsec   &     &  \\
  (1)  &     (2)       &      (3)      &       (4)          &        (5)      &    (6)    &  (7) &  (8)\\
\hline
1........... &  $\ \ \ $7.7555 & +26.4441 & $\ \,$2.1$\,\times\,$10$^{-14}$  & 0.3$\,\times\,$10$^{-14}$  &  26.2 &  7.73 & \\
2........... &  $\ $24.1793 & +20.8143 & $\ \,$1.0$\,\times\,$10$^{-14}$  & 0.2$\,\times\,$10$^{-14}$ &  29.6  & 6.85 & P\\
3........... &  $\ $44.3881 & $\,$-$\,$23.4017 & $\ \,$0.3$\,\times\,$10$^{-14}$ & 0.1$\,\times\,$10$^{-14}$ &  17.7 &  2.91 & \\
4........... &  $\ $46.4686 & +03.8439 & $\ \,$0.8$\,\times\,$10$^{-14}$ & 0.3$\,\times\,$10$^{-14}$ & 21.3 & 3.82 & \\
5........... &  $\ $54.4602 & $\,$-$\,$23.0730 & $\ \,$2.4$\,\times\,$10$^{-14}$  & 0.3$\,\times\,$10$^{-14}$  &  27.2 &  9.52 & \\
6........... &  $\ $73.5050 & $\,$-$\,$03.1444 & $\ \,$2.0$\,\times\,$10$^{-14}$ & 0.3$\,\times\,$10$^{-14}$  &  27.3 &  6.42 & \\
7........... &  $\ $73.6638 & $\,$-$\,$03.1277 & $\ \,$0.7$\,\times\,$10$^{-14}$  & 0.2$\,\times\,$10$^{-14}$   &  16.6 &  3.84 & \\
8........... &  $\ $73.7697 & $\,$-$\,$10.2003 & $\ \,$3.2$\,\times\,$10$^{-14}$  & 0.3$\,\times\,$10$^{-14}$  &  49.3 & 10.95 & \\
9........... &  $\ $73.7895 & $\,$-$\,$10.1752 & $\ \,$1.4$\,\times\,$10$^{-14}$  & 0.2$\,\times\,$10$^{-14}$  &  37.5 &  6.71 & \\
10.......... &  $\ $85.6665 & $\,$-$\,$40.9155 & $\ \,$0.5$\,\times\,$10$^{-14}$  & 0.1$\,\times\,$10$^{-14}$  &  14.7 &  6.63 & P\\
11.......... &  125.0098 & +63.7611 & $\ \,$1.7$\,\times\,$10$^{-14}$  & 0.3$\,\times\,$10$^{-14}$  &  30.2 &  5.79 & \\ 
12.......... &  133.0950 & +51.3794 & $\ \,$2.1$\,\times\,$10$^{-14}$ & 0.2$\,\times\,$10$^{-14}$ &  36.4 & 13.45 & \\ 
13.......... &  133.2558 & +51.2218 & $\ \,$1.2$\,\times\,$10$^{-14}$  & 0.1$\,\times\,$10$^{-14}$ &  26.1 & 11.33 & B \\  
14.......... &  137.5350 & +54.3149 & $\ \,$1.4$\,\times\,$10$^{-14}$  & 0.1$\,\times\,$10$^{-14}$ &  27.4 & 15.45 & \\ 
15.......... &  142.7633 & +79.2219 & $\ \,$3.0$\,\times\,$10$^{-13}$   & 0.1$\,\times\,$10$^{-13}$ & 75.5 & 29.45 & \\
16.......... &  148.6100 & +68.9119 & $\ \,$0.7$\,\times\,$10$^{-14}$  & 0.1$\,\times\,$10$^{-14}$  &  26.5 &  6.37 & B\\
17.......... &  158.8064 & +57.8383 & $\ \,$1.0$\,\times\,$10$^{-14}$  & 0.1$\,\times\,$10$^{-14}$  &  29.3 &  7.74 & \\
18.......... &  158.8564 & +57.8472 & $\ \,$1.2$\,\times\,$10$^{-14}$ & 0.1$\,\times\,$10$^{-14}$  &  29.1 &  8.69 & \\         
19.......... &  164.0525 & $\,$-$\,$03.5845 & $\ \,$5.0$\,\times\,$10$^{-14}$  & 0.2$\,\times\,$10$^{-14}$  &  33.7 & 19.27 & P\\
20.......... &  169.3591 & +07.7274 & $\ \,$2.6$\,\times\,$10$^{-14}$  & 0.3$\,\times\,$10$^{-14}$ &  36.9 &  8.84 & 1\\
21.......... &  169.3746 & +07.7716 & $\ \,$0.9$\,\times\,$10$^{-14}$  & 0.2$\,\times\,$10$^{-14}$   &  26.7 &  4.77 & 2\\
22.......... &  175.0341 & $\,$-$\,$26.5263 & $\ \,$1.5$\,\times\,$10$^{-14}$  & 0.2$\,\times\,$10$^{-14}$  &  32.0 &  8.23 & \\
23.......... &  199.2258 & +29.2394 & $\ \,$2.4$\,\times\,$10$^{-15}$ & 0.5$\,\times\,$10$^{-15}$ &  19.5 &  6.05 & \\
24.......... &  199.3051 & +29.3089 & $\ \,$1.2$\,\times\,$10$^{-15}$ & 0.4$\,\times\,$10$^{-15}$ &  14.9 &  4.14 & \\
25.......... &  204.4194 & +29.4313 & $\ \,$2.0$\,\times\,$10$^{-15}$ & 0.8$\,\times\,$10$^{-15}$ &  17.3 &  3.14 & \\
26.......... &  209.8918 & +62.3164 & $\ \,$3.1$\,\times\,$10$^{-13}$ & 0.6$\,\times\,$10$^{-14}$ &  69.4 & 49.72 & 3 \\
27.......... &  218.1304 & $\,$-$\,$01.1376 & $\ \,$1.4$\,\times\,$10$^{-14}$ & 0.3$\,\times\,$10$^{-14}$ &  26.4 &  5.82 & \\
28.......... &  233.5057 & +23.6009 & $\ \,$1.1$\,\times\,$10$^{-14}$ & 0.1$\,\times\,$10$^{-14}$ &  19.7 &  8.21 & \\
29.......... &  234.9979 & $\,$-$\,$03.0390 & $\ \,$5.7$\,\times\,$10$^{-14}$ & 0.3$\,\times\,$10$^{-14}$ &  51.1 & 19.89 & B\\ 
30.......... &  237.3827 & +21.5499 & $\ \,$3.4$\,\times\,$10$^{-14}$ & 0.3$\,\times\,$10$^{-14}$ &  33.4 & 14.15 & \\ 
31.......... &  245.9976 & +26.6492 & $\ \,$0.5$\,\times\,$10$^{-14}$ & 0.1$\,\times\,$10$^{-14}$ &  14.5 &  4.50 & \\  
32.......... &  249.1558 & +41.1306 & $\ \,$0.8$\,\times\,$10$^{-14}$ & 0.1$\,\times\,$10$^{-14}$ &  25.6 &  9.85 &  \\
33.......... &  255.7827 & +51.5614 & $\ \,$0.9$\,\times\,$10$^{-14}$ & 0.2$\,\times\,$10$^{-14}$ &  29.3 &  5.08 & B \\
34.......... &  310.9874 & +77.1801 & $\ \,$0.7$\,\times\,$10$^{-14}$ & 0.2$\,\times\,$10$^{-14}$& 24.1 & 3.46 & \\
35.......... &  333.2226 & $\,$-$\,$22.1602 & $\ \,$1.1$\,\times\,$10$^{-14}$ & 0.2$\,\times\,$10$^{-14}$ &  23.7 &  6.03 & \\
36.......... &  345.7229 & +08.8118 & $\ \,$1.3$\,\times\,$10$^{-15}$ & 0.4$\,\times\,$10$^{-15}$ &  12.4 &  3.81 & \\
\hline
\hline
\end{tabular}
\end{center}
\vspace{1mm} NOTES - (1) z=0.4 cluster (N. 97 Vikhlinin et al. 1998),
(2) z=0.16 cluster (RIXOS F$258_{-}101$, Mason et al. 2000), (3)
cluster MS 1358.4+6245 at z = 0.328, (P) flux uncertain due to the
presence of a nearby pointlike source, (B) source very close to the
border of the X-ray image.  
\normalsize
\end{table*}

\section{Sky coverage of the survey}
\label{sec:skycoverage}
\begin{figure}[t]
\psfig{figure=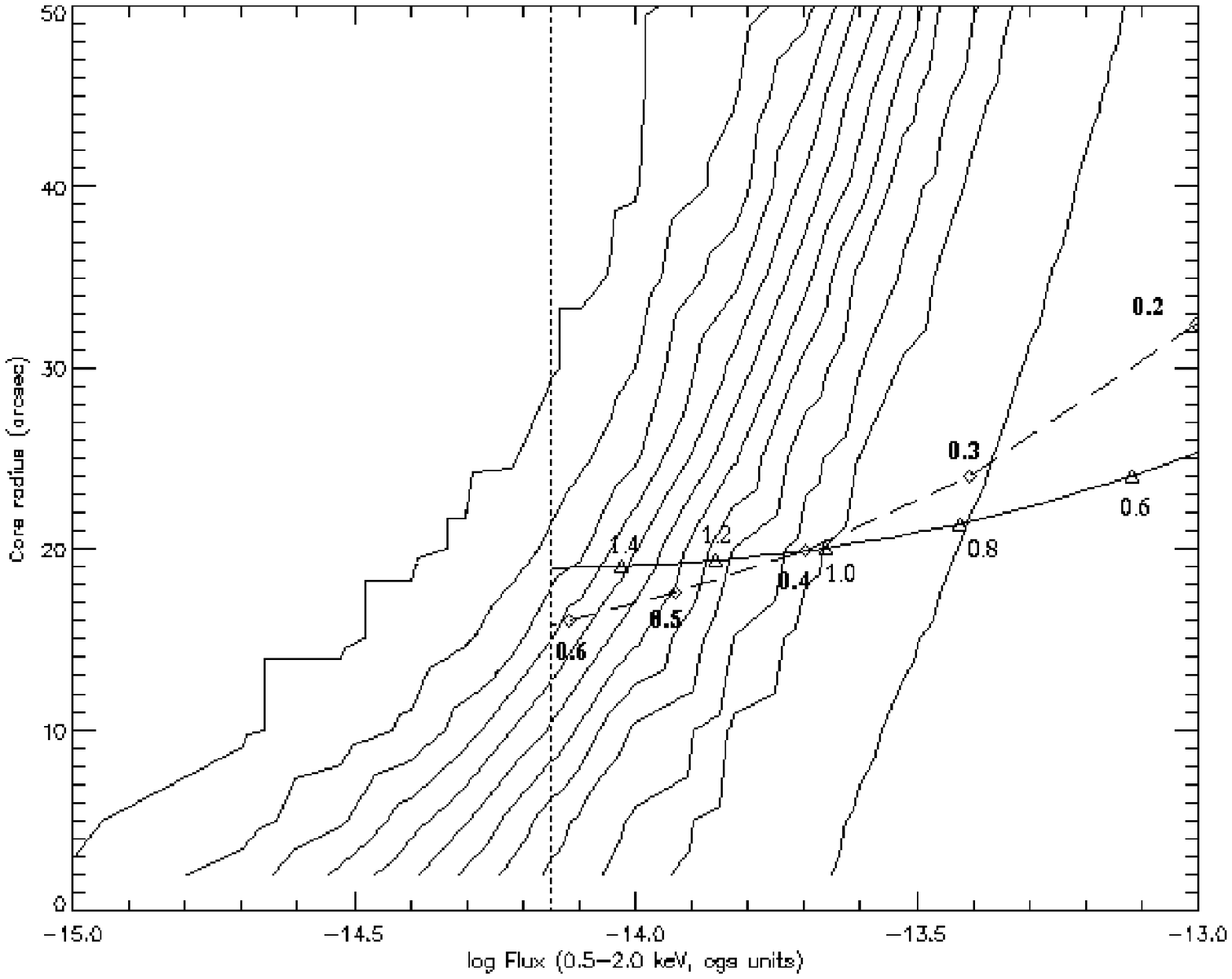,width=9cm,angle=0}
\caption{Sky coverage of the 81 fields used in this survey.  Sky
coverage is plotted as a function of intrinsic flux and intrinsic,
projected core radius (assuming King profile). Contours are drawn at a
percentage of the total survey area. From the left, the first contour
is at a level of 1\%, subsequent solid contours are 10\%, 20\%,...,
90\%, 95\% and 100\%. The two curves at the right-hand side of the
figure are the loci, with redshift, of galaxy groups
($L_{\mathrm{X}}=$ 1 $\times \,10^{43}$ erg s$^{-1}$,
$r_{\mathrm{c}}=100$ kpc, dashed line) and galaxy clusters
($L_{\mathrm{X}}=$ 1 $\times \,10^{44}$ erg s$^{-1}$,
$r_{\mathrm{c}}=150$ kpc, solid line). Redshifts for these systems
increase right to left. The points corresponding to some specific
redshift values are marked and labeled on the two curves. The vertical
dashed line corresponds to the flux value 7 $\times$ 10$^{-15}$ erg
s$^{-1}$ cm$^{-2}$ (see Sect.~\ref{sec:lognlogs}).}
\label{fig:H3892F17.ps}
\end{figure}

As already pointed out in the introduction, the aim of a new X-ray
selected cluster catalogue is twofold: providing a set of targets for
optical/NIR follow-up observations of single interesting objects and
to address global statistical properties of the cluster population
useful for cosmological studies. This second purpose requires the
evalution of the statistical properties of the survey sample (e.g. the
log$\,N$-log$\,S$ and the luminosity functions). To do this, the
knowledge of the effective sky coverage, i.e. the survey area in which
all sources above a given flux are detected, is necessary.\\
Since this survey uses pointed data, each field has a different
detection sensitivity to objects of a given extent and flux, according
to exposure and background. In order to estimate this, VTP's detection
sensitivity can be measured parametrizing a detection criterion based
on a definition of the source signal-to-noise ratio. In particular,
from simulations and real data I find that the criterion for source
detection is approximable as
\begin{equation}
SNR\equiv \frac{n_{\mathrm{\scriptscriptstyle{VTP}}}-n_{\mathrm{bkg}}}{\sqrt{n_{\mathrm{\scriptscriptstyle{VTP}}}}}>3,
\end{equation}
where $n_{\mathrm{\scriptscriptstyle{VTP}}}$ is the total number of
photons (source and background) that lie within the equivalent source
radius
$r_{\mathrm{\scriptscriptstyle{VTP}}}=(A_{\mathrm{\scriptscriptstyle{VTP}}}/\pi)^{1/2}$,
where $A_{\mathrm{\scriptscriptstyle{VTP}}}$ is, as previously
defined, the source area within which the surface brightness exceeds
the VTP surface brightness threshold
$\sigma_{\mathrm{\scriptscriptstyle{VTP}}}$, which is defined by the
background ($\sigma_{\mathrm{bkg}}$) and threshold ($f$) as
$\sigma_{\mathrm{\scriptscriptstyle{VTP}}}=f\,\sigma_{\mathrm{bkg}}$
($f\ge 1.0$).  Assuming again for the sources a King profile as in Eq.
(\ref{eq:kingprofile}), it is possible to compute the SNR for sources
with given fluxes and extents in fields with a given exposure time and
background surface brightness (see Scharf et al. 1997 for the details
of this procedure). Exposure and background information are available
for each of the 81 fields used in this survey, therefore I use them to
integrate the sky coverage over the area of each ACIS field (including
also the PSF variation with off-axis angle) adopting the criterium
introduced above to determine the detection sensitivity.  The final
result shown in Fig.~\ref{fig:H3892F17.ps} is the combined sky
coverage of all fields used in the survey to the limiting threshold
($f=1.0$)\footnote{For calculations in this and in the following
sections cosmic K-corrections from Jones et al. 1998 are used.}.  In
this figure I also plot two curves representing the loci, with varying
redshift, of galaxy groups ($L_{\mathrm{X}}=$ 1 $\times \,10^{43}$ erg
s$^{-1}$, $r_{\mathrm{c}}=100$ kpc, dashed line) and galaxy clusters
($L_{\mathrm{X}}=$ 1 $\times \,10^{44}$ erg s$^{-1}$,
$r_{\mathrm{c}}=150$ kpc, solid line). Redshifts therefore increase
right-to-left along these curves.

\section{Expected redshift distribution of the detected clusters}
\begin{figure}[!t]
\psfig{figure=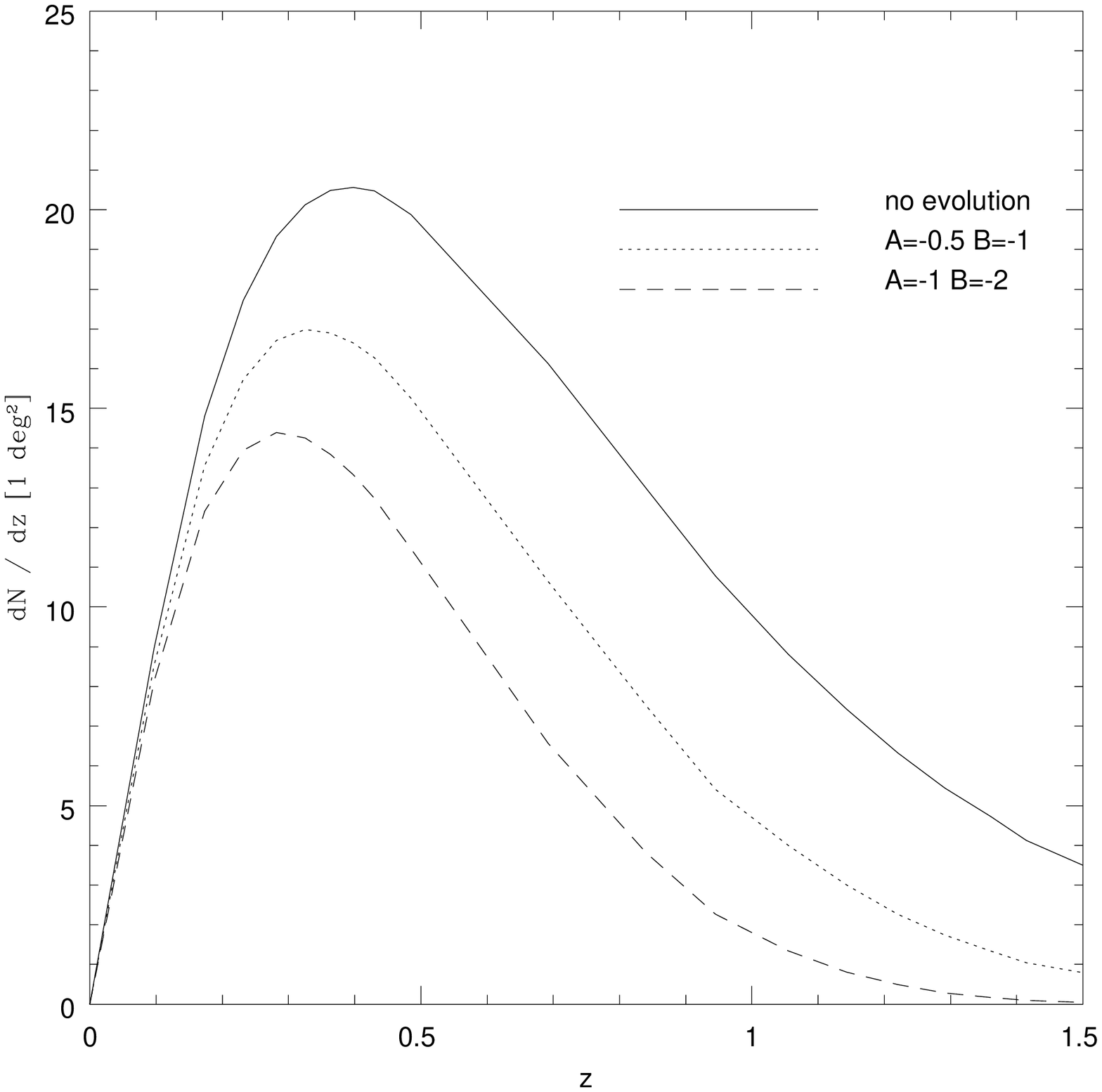,width=9cm,angle=0}
\caption{Expected redshift distribution of the clusters detected in
this survey in the redshift range $0\leq z\leq 1.5$. Three curves
related to three different couples of the XLF evolutionary parameters
(see text) are drawn.}
\label{fig:H3892F18.ps}
\end{figure}
\begin{figure}[t]
\psfig{figure=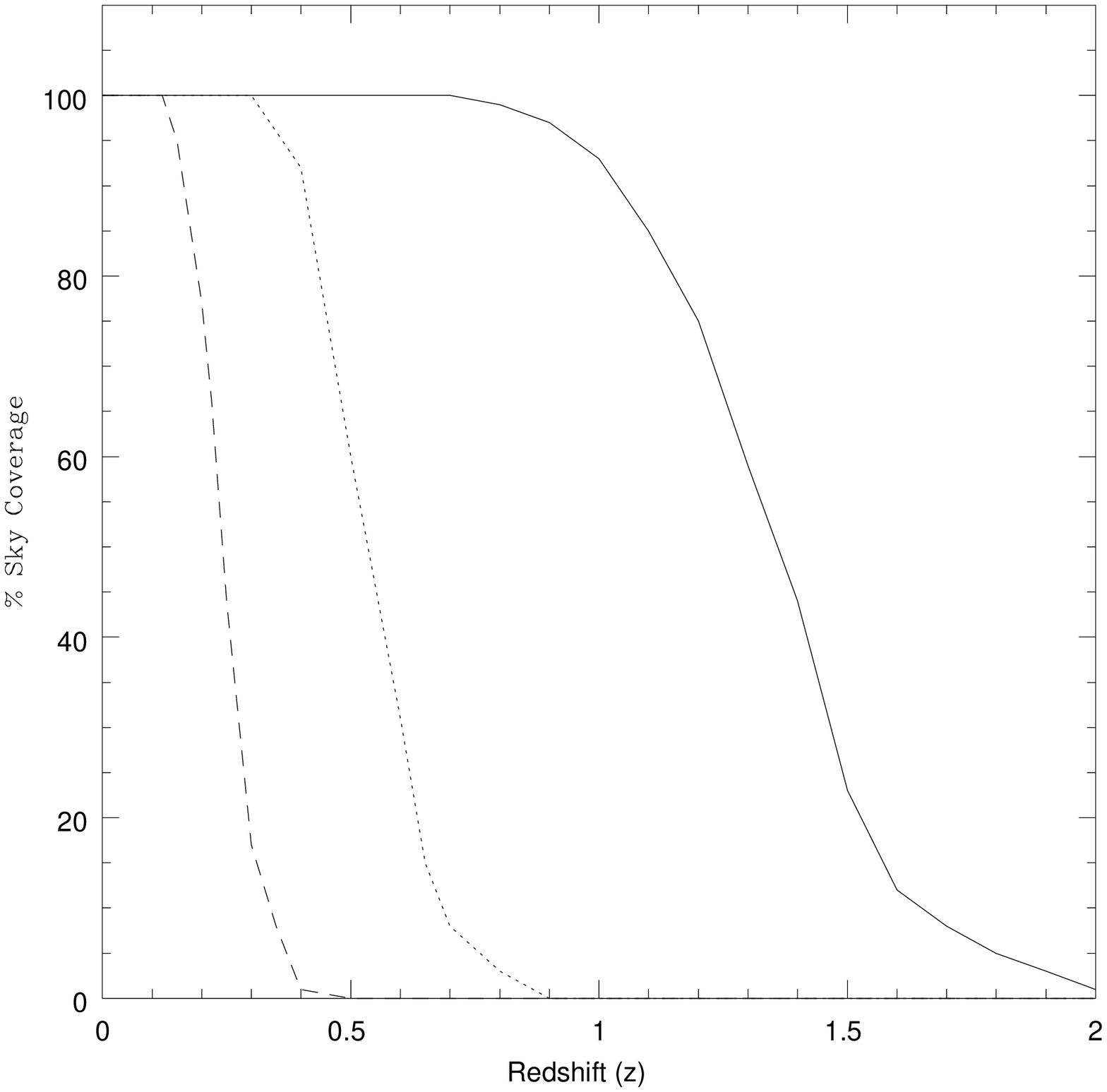,width=9cm,angle=0}
\caption{Sky coverage as a function of redshift for three classes of
objects: elliptical galaxies (with $L_{\mathrm{X}}=$ 1 $\times
\,10^{42}$ erg s$^{-1}$ and effective core radius $r_{\mathrm{c}}=30$ kpc) in
dashed line, groups ($L_{\mathrm{X}}=$ 1 $\times \,10^{43}$ erg
s$^{-1}$, $r_{\mathrm{c}}=100$ kpc) in dotted line, and clusters
($L_{\mathrm{X}}=$ 1 $\times \,10^{44}$ erg s$^{-1}$, $r_{\mathrm{c}}=150$ kpc) in
solid line.}
\label{fig:H3892F19.ps}
\end{figure}
\begin{figure}[t]
\psfig{figure=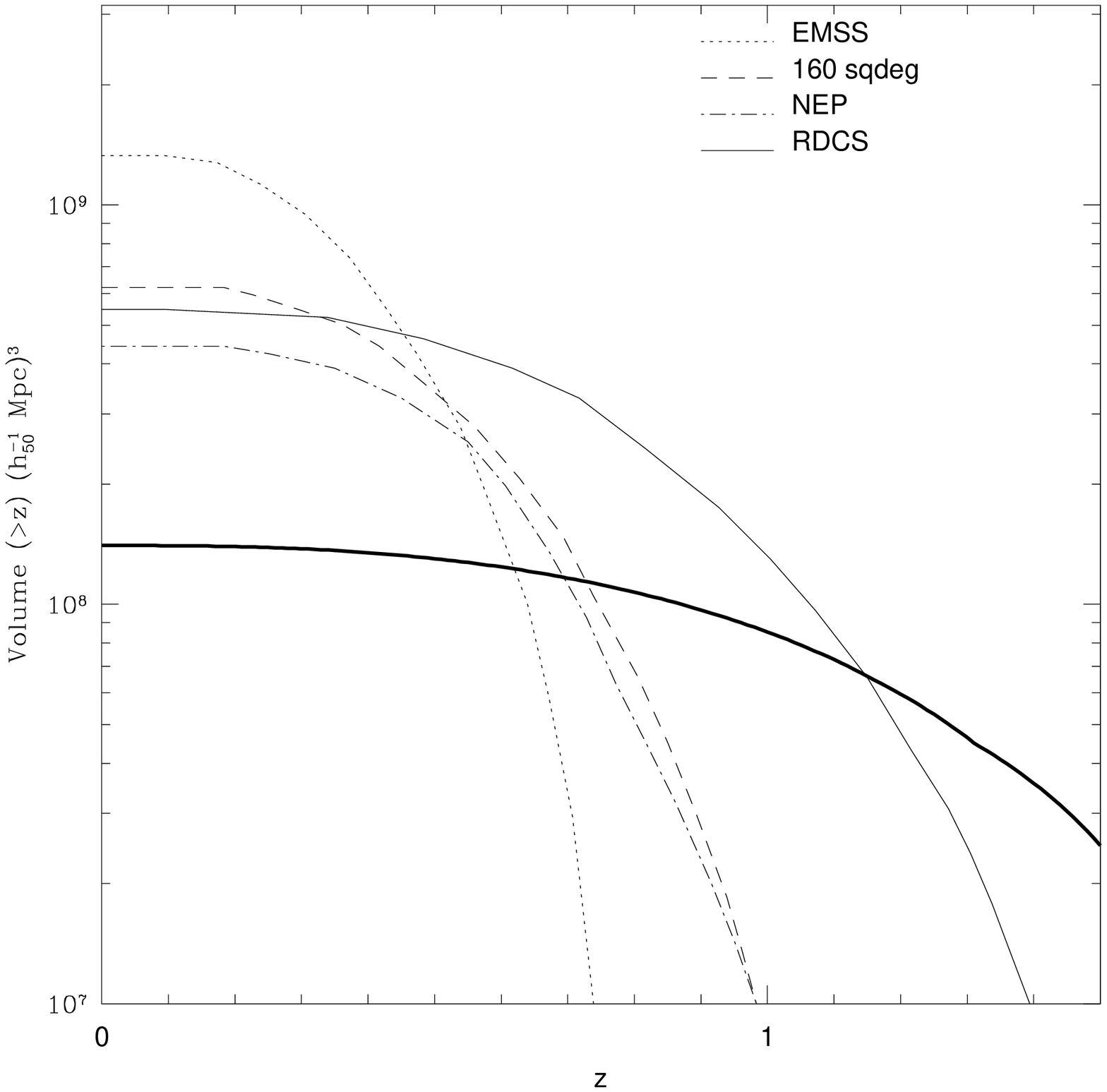,width=9cm,angle=0}
\caption{Search volumes of five X-ray cluster surveys, $V(>z)$, for a
cluster of given X-ray luminosity ($L_{\mathrm{X}}=$ 3 $\times
\,10^{44}\,h_{50}^{-2}$ erg s$^{-1} \simeq L_{\mathrm{X}}^{*}$). The thick
solid line corresponds to the volume probed by this survey (adapted
from Rosati et al. 2002).}
\label{fig:H3892F20.ps}
\end{figure}

An important issue to discuss when compiling a cluster survey is the
estimate of the cluster redshift distribution. This is the comoving
volume per unit redshift and solid angle ($dV/dz\,d\Omega$) times the
comoving density of clusters $n_{\mathrm{com}}$ with luminosities
above the survey detection limit $L_{\mathrm{X}}^{\mathrm{lim}}$. I
write it as
\begin{equation}
\frac{dN}{dz\,d\Omega}=\frac{dV_{\mathrm{com}}}{dz\,d\Omega}n_{\mathrm{com}}=\frac{dV_{\mathrm{com}}}{dz\,d\Omega}\int_{L_{\mathrm{X}}^{\mathrm{lim}}(z)}^{\infty}dL_{\mathrm{X}}\,\phi(L_{\mathrm{X}}),
\end{equation}
where $\phi(L_{\mathrm{X}})$ is the cluster X-ray luminosity function
(XLF), i.e. the function describing the number of clusters per unit
volume with luminosity in the range
$L_{\mathrm{X}}\,\div\,L_{\mathrm{X}}+dL_{\mathrm{X}}$. Now,
$dV_{\mathrm{com}}/dz\,d\Omega$, the comoving volume per unit solid
angle and redshift, is easily computable once a given cosmology is
fixed. $L_{\mathrm{X}}^{\mathrm{lim}}$ is a function of redshift and
depends on the assumed cosmological model, on the shape of the cluster
surface brightness profile and on the exposure times of the
observations. I compute $L_{\mathrm{X}}^{\mathrm{lim}}$ at different
redshifts assuming, as in the previous section, a King profile for
clusters, a typical exposure time of 40 ks (which is close to the
median value of the exposure times of the selected pointings) and
$SNR\ge 3$ as a criterium for cluster detection. For simplicity I also
assume a cluster core radius of 150 kpc, a quite common value
(e.g. Ota \& Mitsuda 2002). Finally I have to integrate the cluster
XLF. Following Rosati et al. (2000), I assume it is an evolving
Schechter function:
\begin{equation}
\phi(L_{\mathrm{X}})=\phi_{\mathrm{0}}\,(1+z)^{A}\,L_{\mathrm{X}}^{-\alpha}\,exp(-L_{\mathrm{X}}/L_{\mathrm{X}}^{*}),
\end{equation}
with $L_{\mathrm{X}}^{*}=L_{\mathrm{X,0}}^{*}(1+z)^{B}$, where A and B
are two constants parametrizing the possible evolution of the XLF with
redshift. For the local XLF, I assume the one derived from the BCS
sample (Ebeling et al. 1997). The results are shown in
Fig.~\ref{fig:H3892F18.ps}. The three curves represent the expected
cluster redshift distribution computed for three different couples of
the evolutionary parameters: $A=-0.5$ and $B=-1$ (dotted line), $A=-1$
and $B=-2$ (dashed line), and $A=B=0$ (no evolution, solid line).
These curves show that the expected cluster redshift distribution has
a maximum at intermediate redshift, at $z\sim 0.3-0.4$.  However, a
significant fraction of the clusters (about 30-40\% for the two
evolutionary models, higher for the no evolution case) are expected to
have redshifts higher than 0.6-0.7, with the possibility of having
clusters at $z>1$. The details of this result are dependent on the
assumptions done above on cluster profiles. However, I find that
changing the cluster profile parameters does not affect the
qualitative behaviour of the cluster redshift distribution.  Of
course, at high redshifts I expect to discover only relatively rich
systems, while at low and intermediate redshift the contribution of
small galaxy groups becomes significant. This is evident from
Fig.~\ref{fig:H3892F19.ps}, where the fractional sky coverage is shown
as a function of redshift for three classes of objects. I have
complete sky coverage for clusters with $L_{\mathrm{X}}=$ 1 $\times
\,10^{44}$ erg s$^{-1}$ and core radius $r_{\mathrm{c}}=150$ kpc out
to a redshift $z\sim 0.8$, and 50\% coverage at a redshift $z\sim
1.3-1.4$. Instead, I have a good coverage for groups (or faint, small
clusters) of $L_{\mathrm{X}}=$ 1 $\times \,10^{43}$ erg s$^{-1}$ out
to $z\sim 0.5$ (50\% coverage). Galaxies have a $\geq 50\%$ coverage
out to $z\sim0.2-0.3$. Therefore, galaxies observed in Chandra fields
are generally low redshift objects. They can be quite easily
recognized in the fields of the Digitized Sky Survey and, eventually,
excluded from the catalogue of extended sources, as explained in
Sect.~\ref{sec:computation}.\\
The knowledge of the sky coverage also allows to compute the volume
that the survey probes above a given redshift $z$, for a given X-ray
luminosity, e.g.  the characteristic luminosity $L_{\mathrm{X}}^{*}$
corresponding to the ``knee'' of the XLF. Rosati et al. (2002) compute
the search volumes of four cluster surveys: EMSS (Gioia et al. 1990),
160 deg$^{2}$ survey (Vikhlinin et al. 1998), NEP survey (Henry et
al. 2001) and RDCS (Rosati et al. 1998). In Fig.~\ref{fig:H3892F20.ps}
I plot the volume probed by these four surveys as a function of
redshift together with the volume probed by this survey.  Since these
surveys cover different solid angles at varying fluxes, they probe
different volumes at increasing redshift and therefore different
ranges in X-ray luminosities at varying redshifts. From the plot we
see that the EMSS is very sensitive to the most luminous systems, but
is able to find only a few clusters at high redshifts due to its
bright flux limit. NEP, 160 deg$^{2}$ and RDCS are much deeper surveys
and discover many clusters at high redshifts.  In particular, the RDCS
pushes the search to the faintest fluxes, providing sensitivity to the
highest redshift systems with characteristic luminosity
$L_{\mathrm{X}}^{*}$ even beyond $z=1$. Instead, the NEP and the 160
deg$^{2}$ surveys do not reach the flux limit of RDCS and probe the
bright end of the XLF out to $z\simeq 1$. This survey (the thick solid
line in the figure) probes a small total volume with respect to the
other surveys bacause of the still limited surveyed solid
angle. Nevertheless, it is superior to NEP and 160 deg$^{2}$ surveys in
probing the Universe at $z\gtrsim 0.8$ and competitive with RDCS for
redshifts $z\gtrsim 1$.

\section{log$\,N$-log$\,S$ relation for the candidate clusters}
\label{sec:lognlogs}
\begin{figure}[t]
\psfig{figure=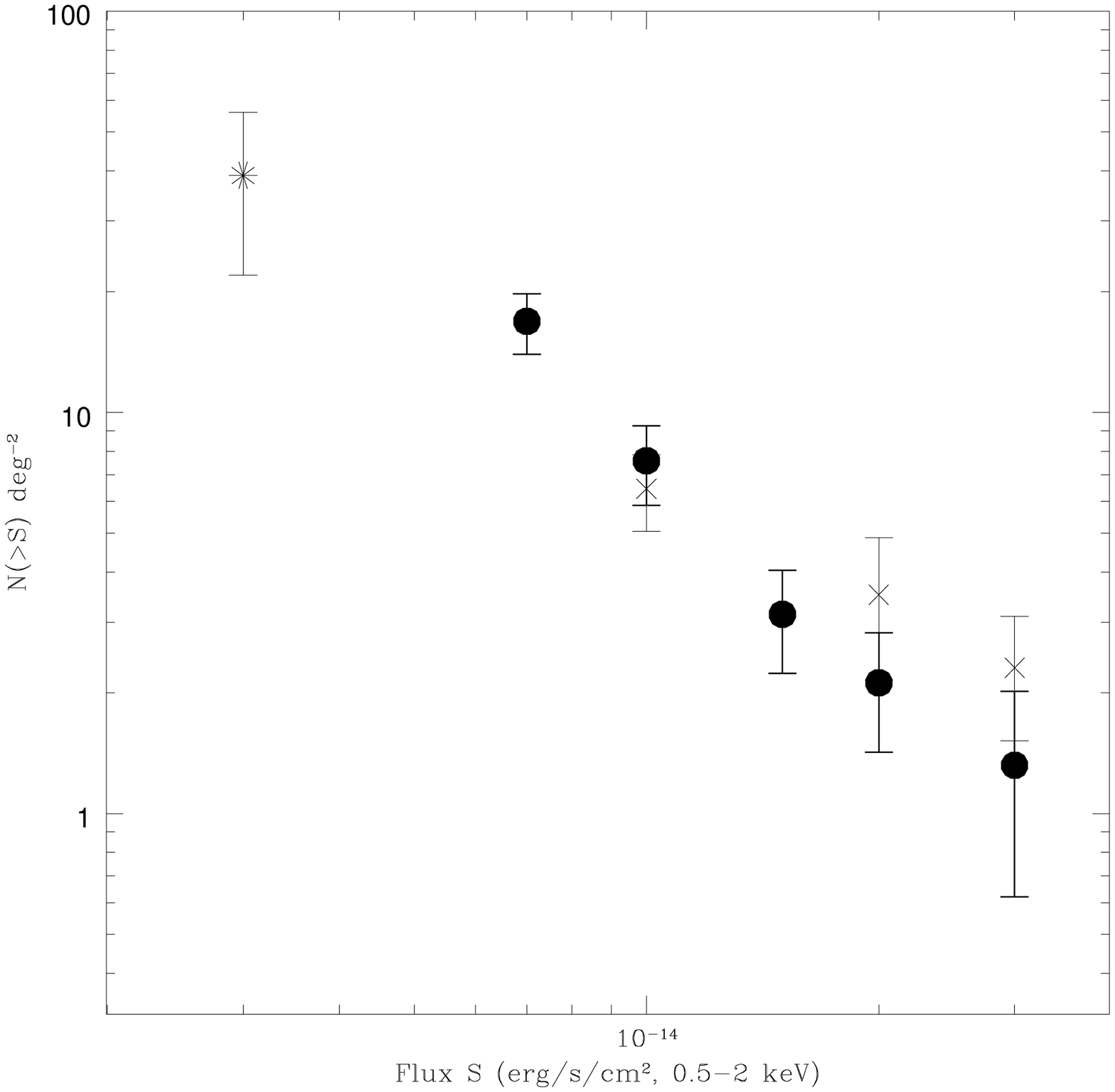,width=9cm,angle=0}
\caption{Cumulative log$\,N$-log$\,S$ function of the candidate
clusters. Counts are shown as filled circles. The error bars represent
the $1\sigma$ uncertainties in the number of detected clusters. Data
from Rosati et al. (1995, crosses) and from McHardy et al. (1998,
asterisk) are also shown for comparison.}
\label{fig:H3892F21.ps}
\end{figure}

The computed sky coverage of the survey is useful to derive the
statistical properties of the cluster population. In particular, it is
interesting to calculate the cumulative number density of the
candidate clusters as a function of flux (the so-called
log$\,N$-log$\,S$ relation) and compare it with previous studies. In
order to perform this calculation, each cluster is added to the
cumulative distribution with the weight equal to the inverse solid
angle corresponding to its measured flux and angular core radius. The
derivative cumulative log$\,N$-log$\,S$ function is shown in
Fig.~\ref{fig:H3892F21.ps}.
Counts are shown as filled circles. The error bars represent the
$1\sigma$ uncertainties in the number of detected clusters. Shown on
abscissa is the total source flux, where ``total flux'' refers to the
flux extrapolated to surface brightnesses below the detection limit
(see Sect.~\ref{sec:correction}). I also show the cluster counts
derived from other two deep surveys: The RDCS (Rosati et al. 1995,
crosses) and an ultra deep UK ROSAT survey (McHardy et al. 1998,
asterisk). The log$\,N$-log$\,S$ relation derived from this survey
spans a flux range from 7 $\times$ 10$^{-15}$ erg s$^{-1}$ cm$^{-2}$
to 3 $\times$ 10$^{-14}$ erg s$^{-1}$ cm$^{-2}$. The lower flux limit
is set at a level in which the sky coverage is about 1 deg$^{2}$ and
all clusters are expected to be detected as extended sources since
they have core radii larger than about 15 arcsec, based on a range of
typical sizes and luminosities (see Fig.~\ref{fig:H3892F17.ps}).
Therefore, I expect this sample be complete to this flux level.
At the bright end, the results are in agreement within errors with the deep 
sample from Rosati et al. (1995).
At the faintest end, the extrapolation of the log$\,N$-log$\,S$ agrees
with the results of McHardy et al. (1998), who identified most of the
X-ray sources, regardless of extent, in their ultradeep ROSAT survey.

\section{Summary}
I present a catalogue of 36 candidate clusters 
detected as extended sources in 81 Chandra ACIS observations covering
an angle of 5.55 deg$^{2}$. To detect these sources, I use a detection
algorithm based on the Voronoi Tessellation and Percolation
technique. This is an efficient method in order to detect low surface
brightness extended sources in a shape independent way. Despite the
small solid angle covered so far, the survey probes a volume at high
redshift ($z\gtrsim 0.8$) comparable or larger than previous Einstein
and ROSAT-based surveys. Moreover, I expect a significative fraction
of the candidates are medium-high redshift ($z > 0.6-0.7$)
systems. This makes these candidate clusters interesting targets for
optical/NIR follow-up observations. In particular, the confirmed distant
clusters will be useful, for instance, to study the properties of
the cluster galaxy population at high redshift (e.g. Gioia et
al. 1999; Stanford et al. 2002). I also present a preliminary
log$\,N$-log$\,S$ relation derived from the catalogue. This relation
shows a general agreement with some deep ROSAT-PSPC based cluster
surveys. As soon as new data will be available in the Chandra archive,
they will be reduced in order to increase the sky area and the volume
probed by the survey and enrich the catalogue compiled so far.

\begin{acknowledgements}
 The data used in this work have been obtained from the Chandra data
 archive at the NASA Chandra X-ray center
 (http://asc.harvard.edu/cda/) and from the On-line Digitized Sky
 Survey server at ESO/ST-ECF Archive
 (http://arch-http.hq.eso.org/dss/dss).\\ 
 I wish to thank Andrea Biviano, Stefano Borgani, Marisa Girardi and
 Massimo Ramella for useful suggestions and discussions. I am
 particularly grateful to Paolo Tozzi for having introduced me in the
 art of X-ray data reduction.\\
 Work partially supported by the Italian Ministry of Education,
 University, and Research (MIUR, grant Progetto Giovani 2001, grant
 COFIN2001028932 "Clusters and groups of galaxies, the interplay of
 dark and baryonic matter"), and by the Italian Space Agency (ASI).
\end{acknowledgements}

{}


\begin{thebibliography}{}

\bibitem{} Arnaud M., Majerowicz S., Lumb D., et al., 2002, \aap, 390, 27

\bibitem{} Bahcall N.A., \& Cen R., 1993, \apj, 407, 49

\bibitem{} Borgani S., \& Guzzo L., 2001, \nat, 409, 39

\bibitem{} Borgani S., Rosati P., Tozzi P., et al., 2001, \apj, 561, 13

\bibitem{} Cavaliere A., \& Fusco Femiano R., 1976, \aap, 49, 137

\bibitem{} Chandra X-ray Center Team, 2001, The Chandra Proposers' Observatory Guide, version 4.0

\bibitem{} Collins C.A., Guzzo L., B\"ohringer H., et al., 2000, \mnras, 319, 939

\bibitem{} Dickey J.M., \& Lockman F.J., 1990, \araa, 28, 215

\bibitem{} Dressler A., Oemler A. Jr., Couch W.J., et al., 1997, \apj, 490, 577 
\bibitem{} Ebeling H., 1993, MPE report 250 (ISSN 0718-0719)

\bibitem{} Ebeling H., \& Wiedenmann G., 1993, \pre, 47, 704

\bibitem{} Ebeling H., Voges W., B\"ohringer H., et al., 1996, \mnras, 281, 799

\bibitem{} Ebeling H., Edge A.C, Fabian A.C., et al. 1997, \apj, 479, 101

\bibitem{} Ellingson E., Lin H., Yee H.K.C., \& Carlberg R.G., 2001, \apj, 547, 609 
\bibitem{} Fruscione A., \& Siemiginowska A., 2000, Chandra News, 7, 4

\bibitem{} Gioia I.M., Henry J.P., Maccacaro T., et al., 1990, \apjl, 356, 35 

\bibitem{} Gioia I.M., Henry J.P., Mullis C.R., \& Ebeling H., 1999, \aj, 117, 2608

\bibitem{} Girardi M., Borgani S., Giuricin G., Mardirossian F., \& Mezzetti M., 1998, \apj, 506, 45

\bibitem{} Gonzalez A.H., Zaritsky D., Dalcanton J.J., \& Nelson A., 2001, \apjs, 137, 117

\bibitem{} Gonzalez A.H., Zaritsky D., \& Wechsler R.H., 2002, \apj, 571, 129

\bibitem{} Henry J.P., Gioia I.M., Mullis C.R., et al., 2001, \apjl, 553, 109

\bibitem{} Holden B.P., Stanford S.A., Squires G.K., et al., 2002, \aj, 124, 33

\bibitem{} Jones C., \& Forman W., 1984, \apj, 276, 38

\bibitem{} Jones L.R., Scharf C., Ebeling H., et al., 1998, \apj, 495, 100

\bibitem{} Jorgensen I., Hjorth J., Franx M., \& van Dokkum P.G., 1997, \baas, 29, 780

\bibitem{} Kelson D.D., van Dokkum P.G., Franx M., Illingworth G.D., \& Fabricant D., 1997, \apj, 478, 13

\bibitem{} Kiang T., 1966, \zap, 64, 433

\bibitem{} King I.R., 1962, \aj , 67, 471

\bibitem{} Mason K.O., Carrera F.J., Hasinger G., et al., 2000, \mnras, 311, 456

\bibitem{} Mauskopf P.D., Ade P.A.R., Allen S.W., et al., 2000, \apj, 538, 505

\bibitem{} McHardy I.M., Jones L.R., Merrifield M.R., et al., 1998, \mnras, 295, 641

\bibitem{} McLean B.J., Greene G.R., Lattanzi M.G., \& Pirenne B., 2000, \pasp, 216, 145

\bibitem{} Nelson A.E., Gonzalez A.H., Zaritsky D., \& Dalcanton J.J., 2001, \apj, 563, 629

\bibitem{} Nelson A.E., Gonzalez A.H., Zaritsky D., \& Dalcanton J.J., 2002, \apj, 566, 103

\bibitem{} Nichol R.C., Collins C.A., Guzzo L., \& Lumsden S.L., 1992, \mnras, 255, 21

\bibitem{} Nonino M., Bertin E., da Costa L., et al., 1999, \aaps, 137, 51

\bibitem{} Ota N., \& Mitsuda K., 2002, \apj, 567, 23

\bibitem{} Pointecouteau E., Hattori M., Neumann D., et al., 2002, \aap, 387, 56

\bibitem{} Postman M., Lubin L.M., Gunn J.E., et al., 1996, \aj, 111, 615

\bibitem{} Ramella M., Boschin W., Fadda D., \& Nonino M., 2001, \aap, 368, 3, 776

\bibitem{} Raymond J.C., \& Smith B.W., 1977, \apjs, 35, 419

\bibitem{} Reese E.D., Mohr J.J., Carlstrom J.E., et al., 2000, \apj, 533, 38

\bibitem{} Reiprich T.H., \& B\"ohringer H., 1999, Astr. Nachr., 320, 296 

\bibitem{} Reiprich T.H., \& B\"ohringer H., 2002, \apj, 567, 716

\bibitem{} Rosati P., Della Ceca R., Burg R., Norman C., \& Giacconi R., 1995, \apj, 445, 11

\bibitem{} Rosati P., Della Ceca R., Norman C., \& Giacconi R., 1998, \apjl, 492, 21 

\bibitem{} Rosati P., Borgani S., Della Ceca R., et al., 2000, Proceedings of the Workshop ``Large Scale Structure in the X-ray Universe'', 1999, Santorini (Greece)

\bibitem{} Rosati P., Borgani S., \& Norman C., 2002, \araa, 40, 539

\bibitem{} Scharf C.A., Jones L.R., Ebeling H., et al., 1997, \apj, 477, 79

\bibitem{} Scharf C.A., 2002, \apj, 572, 157

\bibitem{} Schuecker P., B\"ohringer H., Guzzo L., et al., 2001, \aap, 368, 86

\bibitem{} Stanford S.A., Holden B.P., Rosati P., et al., 2002, \aj, 123, 619

\bibitem{} Sunyaev R.A., \& Zeldovich Y.B., 1972, Comments Astrophys. Space Phys., 4, 173 

\bibitem{} van Dokkum P.G., \& Franx M., 2001, \apj, 553, 90

\bibitem{} Vikhlinin A., McNamara B.R., Forman W., et al., 1998, \apj, 502, 558

\bibitem{} Voronoi G., 1908, Journal f\"ur Reine und Angewandte Mathematik, 134, 198

\bibitem{} Wise M., 1997, Chandra News (published as AXAF News), 5, 22

\end{thebibliography}
\end{document}